\documentclass{aa}  
\usepackage{graphicx}
\usepackage{txfonts}
\newcommand{\di}{\mbox{d}}

\begin{document} 

 \title{Ab initio simulations of neutron star oblique electrosphere with realistic neutron star parameters}


 \author{F. Mottez
          \inst{1}
          }

   \institute{Laboratoire Univers et Théories, Observatoire de Paris, CNRS, Université PSL, Université Paris Cité, F-92190 Meudon, France\\
              \email{fabrice.mottez@obspm.fr}           
             }
   \titlerunning{Oblique electrospheres}

 
  \abstract
   {Electrospheres are environments with the same origin as pulsars, associated with a highly magnetized rotating neutron star. In pulsars, a cascade of electron-positron pair creation enriches the plasma. The plasma surrounding an electrosphere consists only of particles escaped from the neutron star surface. Electrospheres with a magnetic axis aligned with the rotation axis have been well described for decades. Models of electrospheres with oblique magnetic axis relatively to the rotation axis resisted to most theoretical investigations. Some electrospheres and pulsars have been simulated with particle-in-cell codes, but the numerical constraints did not allow the use of realistic neutron star parameters. }
   {We aimed at developing a numerical simulation code optimized for the understanding the physics of electrospheres and pulsar, with realistic neutron star parameters. As a first step, presented in this paper, we focused on the simulation of oblique electrospheres with realistic physical parameters.  }
   {A specific code is developed for the computation of stationary solutions. The resolution of the Maxwell's equations is based on spectral methods. Particle motion include their finite inertia. No hypothesis is made relative to a force-free behavior of the electrospheric plasma. The numerical code is  called \textit{Pulsar ARoMa}, for Pulsar Asymmetric Rotating Magnetosphere.}
   {Various numerical simulation were conducted for neutron star realistic parameters. We find that oblique electrospheres possess the same global structure as for aligned force-free electrosphere, with two domes of electrons and a torus of positively charged particles. The domes are not centered on the magnetic axis, and they are asymmetric.  Yet, the solutions do not exhibit a force-free behavior. }
   {The simulations performed with the code \textit{Pulsar ARoMa} require modest resources and little computing time. This code will be upgraded for more ambitious investigations on pulsar physics. }

   \keywords{ --neutron stars, electrospheres, numerical methods, magnetic fields              
               }

   \maketitle
%
\nolinenumbers

\section{Introduction}

Since their discovery  by \cite{Hewish_1969}, considerable progress has been made in modeling pulsars.  However, considerable progress is also needed to really understand the physical mechanisms governing these highly energetic objects.
In this paper, a step forward towards this goal is presented, based on the writing of a new numerical simulation code, named \textit{Pulsar ARoMa} for "Pulsar Asymmetric Rotating Magnetosphere", allowing to solve the fundamental equations governing the electrodynamics of pulsars, in a regime allowing the use of realistic parameters. 

As soon as pulsars were recognized as manifestations of neutron stars environments, the problem could be posed physically in a very simple way as that of a magnetized, electrically conductive, rapidly rotating sphere in an initially empty medium. Generally, the magnetic field at the surface of the star is considered as dipolar, but multipolar components are also possible. The symmetry axis of the magnetic field can make any angle with the rotation axis of the neutron star (oblique pulsars), but for computational ease, solutions with a dipole aligned with the rotation axis (aligned pulsars) have been given special attention. 

The first step in solving the pulsar problem was the calculation of the electromagnetic field into the vacuum extending around the neutron star. For an oblique dipole, a previous computation by \cite{Deutsch_1955} showed that the solution is close to the dipole solution in the vicinity of the star, but, extends beyond the light cylinder in the form of a wave propagating at the speed of light, whose wave front is deployed along an Archimedes spiral. This wave carries a Poynting flux that corresponds in part with the loss of rotational energy of the pulsar \citep{Gold68,Pacini67}. A vacuum solution with multipolar magnetic field was developed more recently \citep{Mottez_2015c, Petri_2015_multipole}. An important element of the vacuum solution is the fact that the rotating neutron star behaves like a unipolar inductor, generating a large-amplitude electric field at the star's surface.

However, it has been shown in \citet{Goldreich69} that the surface electric field repels electrons and ions from the star's surface and accelerates them to very high energies. As a result, the NS is embedded in a plasma, not in an empty region.
\citet{Goldreich69} applied the concept of unipolar inductor to neutron stars, assuming that the plasma near the star was rotating with it. They then developed the fundamental concept of corotation space charge of the plasma, deduced from the divergence of the corotation electric field. Unlike most astrophysical objects, this charge density is high in the vicinity of neutron stars, due to their high rotation rate and magnetic field.
They thus showed that there is a plasma-filled non neutral magnetosphere around neutron stars. The concept of corotation charge cannot be applied at locations near and beyond the light cylinder, but it lays the foundation for understanding the internal boundary conditions of the pulsar, i.e. near its surface.

The electrons extracted from the neutron star surface are  called the primary electrons. They follow a curved trajectory, the curvature being those of the neutron star magnetic field lines (with a curvature radius $\sim 100 $km). Because of the high electric fields above the surface, the primary electron can reach Lorentz factors of several millions, and they emit gamma ray photons. \citet{Sturrock_1970} showed that their interaction with the magnetic field of the neutron star generates energetic pairs of electrons and positrons which, in turn, emit gamma rays and other pairs, thus causing a cascade of pair creation. Pairs can also be created by gamma ray photons interacting with thermal (X-ray) photons emitted by the neutron star.
\cite{Timokhin_2019} made an evaluation of the cascade efficiency  with the help of a local (1D) semi-analytical model of the acceleration region. In highly magnetized pulsars, the efficiency of the cascades can reach a factor of a few $10^5$ pairs created for each primary electron.

All these works highlight important facts about pulsars. But the construction of a global picture remained to be done.  Many works have been devoted to this difficult task. Since the early 2000', most of them are based on numerical simulations. They are reviewed in \citet{Cerutti_2017}. In a seminal work, \cite{Krause_1985} showed that a neutron star magnetosphere without pair creation is formed by two domes of electrons above the polar caps and of a belt of positively charged particles near the magnetic equator. This is called an electrosphere. With only one exception \citep{McDonald_2009}, electrospheres were described only for an aligned magnetic field. 

\cite{Higgins_1997} developed a radiation model for the high-energy emission from pulsars considering charged particle motion in the fields of a spinning, highly magnetized and conducting sphere in vacuum. The electromagnetic field corresponded to the \cite{Deutsch_1955} vacuum solution, and there was no pair creation. The radiation emitted by curvature emission was summed to generate light curves. The calculation was simplified analytically, but something equivalent to the electrosphere appeared. 

In this paper, models of electrospheres are presented. They are based on numerical simulation. They  have a few characteristics that are not met simultaneously in the literature, as far as I know. The bad new is that the models presented in this study are stationary. They do not permit analysis of the possible development of plasma instabilities. The good news are the following. The models include particles with finite inertia. The particles have real masses and real charges. There are no reduced variables, ordinary physical units can be used to characterize a set of simulation parameters, and the model include realistic neutron star radius, rotation rates and magnetic field amplitudes.  The magnetic field emerging from the neutron star is dipolar, but this choice is simply done for the sake of simplicity. There is no \textit{a priori} hypothesis about force-free region. We suppose a zero work function for the extraction of particles from the stellar surface. A large set of models of oblique electrospheres is presented. 
 
This step is intermediate, it does not allow yet to simulate the complete pulsar dynamics, because the version presented here does not take into account the radiative phenomena and the pair creation phenomenon. It allows however to simulate solutions belonging to the electrosphere family, that, from an astronomical point of view, can represent "dead pulsars", beyond the death line (or the death valley) of the $P-\dot P$ diagram \citep{Beskin_2022}.

The units used in the simulations are ordinary CGS units, unless stated otherwise. We frequently use the adjectives "parallel" and "perpendicular" to characterize vectors. This is relatively to the direction of the local magnetic field $\vec{B}$. Of course, these adjectives only make makes sense when $B \ne 0$, otherwise, we don't use them.

\section{Particle inertia and electrospheres}

Before explaining the method used in our numerical simulations, we discuss the role of particle inertia in electrospheres. 
 
A large class of electrosphere models is based on hypothesis introduced by \cite{Krause_1985} of particles without inertia. If such particles encounter a finite parallel electric field, they reach a velocity $c$ and talking about their Lorentz factor is meaningless. Therefore, any equilibrium involving such particles implies force-free conditions $\vec E \cdot \vec B=0$, in any place containing a plasma. Non force-free electromagnetic fields are possible only in vacuum gaps, i.e. regions with no particle. There are two practical consequences regarding electrospheres. First, the dome containing electrons, and the torus containing ions or positrons are force-free. Second, the inner boundary conditions, just above the surface of the star are also force-free. 

An interesting consequence of this last condition is that there is no real connexion between the star and its electrosphere, and the equations describing this family of models of electrospheres show that the total electric charge of the electrosphere is a free parameter. 

\cite{Smith_2001} conducted numerical simulations of aligned electrospheres with particles without inertia. They showed solutions for a large range of magnetospheric global charges, as stated by \cite{Krause_1985}. Their solutions for typical pulsar parameters were all encompassed within the light cylinder, but the authors remarked that,  if there were a loss in the system, it would not lead to replacement by particles escaping from the star. 

\cite{Petri_2002} independently developed  a semi-analytical theory of pulsar aligned electrospheres, starting from the same hypothesis. They reached the same conclusions. They also showed that  the geometric and kinematic structure of the electrosphere is uniquely determined by the total charge of the system, and that the condition for a cascade of electron-positron pair creations are not met (except maybe for a millisecond pulsar). A practical consequence of this last conclusion, compatible with up do date radio astronomical observations, is that aligned electrospheres can exist, but aligned pulsars cannot exist.

In the present study, particles have a finite inertia. They are governed by a sum of the Lorentz electromagnetic force, and a reaction force caused by the radiations that they emit. 
Because of their inertia, particles can be accelerated by finite parallel electric fields without reaching infinite Lorentz factors. Therefore, there is no need to settle force-free conditions above the surface of the star. Hence, the electrosphere is not isolated from the star, and a self consistent electrospheric total charge can be established. It is not an \textit{ad hoc} parameter as is in force-free electrospheres.   

Our simulations contain force-free zones. Particles don't necessarily find an equilibrium position within them, they can oscillate around them, sometimes, as we shall see, with a spatial amplitude that can encompass very large regions.

\cite{Philippov_2014} simulated one aligned electrosphere with a particle in cell (PIC) code. In order to minimize computational expense, the simulation was run with a star having a high rotation rate, resulting in a light cylinder distance $R_{LC}$ of only 2.6 star radii $R_*$. As in force-free models, the electrosphere contains a dome of positrons above each poles, and an equatorial belt of electrons. (The magnetic field axis being opposite to the rotation axis, the signs of electric charges in these two regions are exchanged.)   They showed that the current flow is insufficient to appreciably modify the dipole magnetic field structure or cause pulsar spin-down. Very little is said about the global electric charge of the electrosphere, or about the mapping of the parallel electric field, so it is difficult to compare this simulation with those involving the force-free approximation. 

\cite{McDonald_2009} made the only simulations of oblique electrospheres. A PIC code was used, and particles had a finite inertia. The surface of the star was initially not force-free, but they considered that the convergence toward an electrosphere solution was reached when, at the stellar surface, $\vec{E} \cdot \vec{B}$ approached a null value. Nothing is said in their paper about the particle injection process from the stellar surface into its environment : what were the particles energies, what was the flux of injected particles, etc. There are probably several ways of injecting particles, and they probably have an influence on the properties, and even the existence, of the electrosphere. 

In the present work, we don't impose any force-free condition. We consider that right above the surface of the star, particles can escape at high energies owing to the parallel electric field allowed in our simulations. This does not imply necessarily that  a net flux of particles leaves each portion of the surface. Indeed, particles can go back toward the surface and reach it. (The same thing was possible in \cite{McDonald_2009}, but only before an equilibrium was reached.). The main difference is that the particles coming back into the star are decelerated, while the same particles living the surface are accelerated. Therefore, with finite particle inertia, static equilibrium can be reached through a dynamic process involving particles quitting the surface, and returning back to it (not necessarily in the same place), without the necessity of a null parallel field above the stellar surface. 

\section{Method}

\subsection{General principles}

The main data of a simulation consist of an electromagnetic field recorded on a 3D grid, a charge density and a current density recorded on the same 3D grid, a particle distribution recorded on a 4D phase space grid,  and a series of pseudo-particle trajectories. 

A simulation consists of a series of iterations during which all simulation data are re-evaluated. Since the complete particle trajectories are calculated during an iteration, and since we are looking for a stationary solution, not a time-dependent one, an iteration does not correspond to a time interval. It corresponds to an undefined period of time. An iteration is (ideally) a step in a process of convergence from initial conditions that are not a solution of the pulsar problem to a time-independent solution. The term "ideally" in brackets simply underlines the fact that convergence is not guaranteed.
  
The elements of the pseudo-particle trajectories consist of their contribution $f$ to the phase-space distribution, a position $\vec{x}_i$, a Lorentz factor $\gamma_i$ and a velocity $\vec{v}_i$ at times $t_i$ elapsed since the beginning of the trajectory ($t_0=0$). For practical reasons, a particle trajectory contains a maximum number $P_{max}$ of elements/time steps $i$. 
In most simulations presented here, $P_{max}=10,000$.

The pseudo-particle trajectories obey the equation of motion
\begin{equation} 
\dfrac{\di }{\di t}(\gamma \vec{v})=\frac{q}{m} (\vec{E}+\vec{E}_R+ \vec{v} \times \vec{B}) \label{eq_mouvement_lent_fondamentale}
\end{equation}
where $\vec v$ is the pseudo-particle velocity, $\gamma$ is its Lorentz factor, $q$ and $m$ its electric charge and mass. The velocity is a 3D vector. No particular assumptions are made about the value of its parallel or perpendicular component. The vectors $\vec{E}$ and $\vec{B}$ represent the local electric and magnetic fields. The pseudo electric field $\vec{E}_R$ represents a counter reaction force caused by the particle radiation. More details are given in section \ref{sec_particle_transport}. 

Generally, a particle trajectory is stopped when it quits the simulation domain, either falling onto the star (inner boundary), or when it crosses the outer boundary, at a large distance from the star.  A particle represents a fraction of the particle distribution. When a pseudo-particle ends its trajectory ($i=P_{max}$) and is still in the simulation domain, the last element of the trajectory is stored into a 4D phase space density, and becomes the starting point of a new trajectory that will be computed at the next iteration. Ideally, the phase space should have 6 dimensions representing the positions $\vec{x}$ and the momentum. Actually, the particles in the vicinity of a neutron star have a very high energy. Consequently, they are strong emitters of synchrotron radiation as soon as they have a finite perpendicular momentum. Therefore, the perpendicular momentum quickly tends to zero. We assume that this observation that is true in general, is systematically true, this is why the particle is stored in a 4D phase space whose components represent the 3 dimensions of space, and the parallel momentum. 

The positions in space $\vec{x}_i$ occupied by the different elements of a trajectory are used to calculate the contribution of the pseudo-particle to the charge density $\rho(\vec{x})$ and the current density $\vec{J(\vec{x})}$, whose values are defined on each element of the 3D grid. 

The charge and current densities  on the grid are the source terms of the Maxwell's equations. Once a new electromagnetic field is computed onto the 3D grid, a new iteration can take place, unless we consider that the simulation code has reached a stationary solution. 

The convergence toward a stationary solution is checked "visually", and through an evaluation the total electric charge outside the star. Its variations must converge toward a constant value. Generally, when convergence is reached, the residual amplitude of the  fluctuations of the total charge density of the electrosphere is of the order of one percent. 
In the present paper,  convergence was reached after 3 or 4 iterations, and most simulations  were conducted with 10 or 15 iterations, in order to ensure the numerical stability of the solution. 

\subsection{Simulation grid} \label{sec_grid}
The grid on which the electromagnetic field, the charge density, the current density and the particle distributions are computed is a lattice of lines of constant spherical coordinates. The origin of coordinates is at the stellar center. The principal axis of these coordinates is the stellar rotation axis. There are $N_{angles}$ lines of equally spaced planes of fixed azimuth $\phi_i$, and the same number of cones of equally spaced co-latitude $\theta_i$ ($i \in [1,N_{angles}]$). This grid is appropriate for the resolution of the Maxwell's equations with the use of spherical harmonic functions. Unless stated otherwise, $N_{angles}=32$.

The radial coordinate $r$ is associated with spheres of radius $r_j$. The sphere indexed by a number zero has a radius $r_0$ equal to the star's radius $R_*$. The increments $r_{j+1} - r_j$ are not uniform, because a high resolution (meters or centimeters) is needed above the stellar surface, and a lower resolution is permitted at larger distances. 
The simulation grid is a sphere containing $N_{d}$ domains connected to each other, each domain containing $N_{nodes}$ Gauss-Lobatto nodes defined as $x_j=\cos ({\pi j}/{N_{nodes}})$, that are appropriate with a decomposition of the electromagnetic field  (for the $r$ variable) in Chebychev polynomials \citep{Grandclement_2009}.
The $N_{d}$ domains do not have the same size. 
In most simulations of the present study,   
the first domain, close to the star, has a size of $10^{-4} \times R_*$. After \citet{Riley_Nicer_2021}, a star's radius $R_*=12$ km is set. The first domain has a size of 1.2 m. The distance between the two first nodes of this domain is $1.15$ cm. There are $N_{d}=11$ domains and  $N_{nodes}=16$ nodes per domain. The size of the second domain is 12 m, then 120 m, 1.2km,  and the following domains are twice as large as the previous ones. Therefore, the total radius of the simulation domain is $1700$ km$ =142 R_*$. The simulation domain contains the light cylinder. 


Particle distributions for each species $s$ are also stored on a grid representing a reduced 4D phase space. 
The first 3 coordinates are the space coordinates $(r,\theta,\phi)$, the last is the absolute value of the parallel impulsion normalized to the particle mass $U=p_\parallel / m_s = \gamma v_\parallel$,
 completed by the sign $s$ of $\vec{p}_\parallel$ relative to the radial direction (negative if the particles goes in the direction of the star). In the present version of the code, the parallel impulsion cells are defined in the same way as the $r$ cells, as a series of $N_{dp}$ domains containing $N_{nodes p}$ that are Gauss Lobatto nodes. Actually, this is for "historical reasons" since spectral methods are finally not used to deal with the particle impulsion. This could be changed.

\subsection{Electromagnetic field solver} \label{sec_field_solver}
As in \cite{Petri_2013}, we use a spectral method to solve the Maxwell's equations, where all the fields are decomposed on a basis of spherical harmonic functions of $\theta$ and $\phi$, and, for each domain, on a basis of Chebychev functions of $r$. A description of this method is given in \citep{Grandclement_2009}, and useful additions to this document are given in the appendix to \cite{Petri_2013}. The main differences with the present work are that \cite{Petri_2013} solves a general relativistic version of Maxwell's equations (for a given metric), whereas we only consider special relativity. In \cite{Petri_2013}, only one simulation domain for the Chebychev functions in $r$ is used, while $N_{d} >1$ domains are used in the present work. 

\subsection{Motion of pseudo particles} \label{sec_motion}
The equation of motion \ref{eq_mouvement_lent_fondamentale} is solved using a relativistic version of an algorithm that solves the motion of the guiding center  \citep{Mottez_1998,mottez_2008_a}. The implementation of special relativity is derived form \cite{Vay_2008}. This method actually provides the guiding center motion of the particle if the time step $\Delta t$ is large compared to the particle gyrogrequency $\Omega_s$, i.e. if $\Delta t \Omega_s >>1$. In the reverse case, it provides the full motion of the particle. 
The highly inhomogeneous environment of a neutron star imposes a very low time step $\Delta t$ near the stellar surface and in acceleration regions; but it would allow higher values in other regions of the electrosphere. In order to increase the computational efficiency of the simulation, the implementation of a method of computation of pseudo-particle trajectories with variable time steps was set and validated. 

In the simulations presented in the following sections, because of the very high electric fields near the star's surface where particles are emitted, we start with a time step  $c \Delta t <0.1$ mm. At greater distances from the star, the algorithm provides a time step up to $c \Delta t \sim  $ 10 m.

This algorithm, its implementation and testing can be useful for other applications and will be presented in a future article, because their field of application goes beyond pulsars and astrophysical applications.

\subsection{Radiative losses by curvature radiation} \label{sec_radiative}
The particles are accelerated to ultra-relativistic energies, and move in a magnetized environment: radiative losses due to the acceleration are very important. In these conditions, loss associated with synchrotron emission annihilate the high frequency motion perpendicular to the local magnetic field. Therefore, their pitch angle is zero. Synchrotron radiation losses are important only when new particles are created. But curvature radiation, caused by the curvature of the particles trajectories along magnetic field lines, cannot be neglected. 
The power loss is (in Gaussian units)
\begin{equation} \label{eq_power_curvature}
P_c=\frac{2}{3} {q^2 c \gamma^4 C_m^2},
\end{equation}
where $C_m$ is the curvature of their trajectory, which we equate with the curvature of local magnetic field lines. 
This power can be likened to the work (per time unit) of a force caused by an electric field $qv \vec{E}_R \sim cq \vec{E}_R$ that is anti-parallel to the particle velocity. The electric field $\vec{E}_R$ is the one appearing in Eq. \ref{eq_mouvement_lent_fondamentale}.

Since the magnetic field is known for each iteration, computing $C_m$ is straightforward. But, with highly-relativistic particles, hence, with large Lorentz factors, the $\gamma^4$ factor in the definition of $\vec{E}_R$ is the potential source of a numerical runaway instability, with  $\gamma$ increasing in a non physical way. Actually, various test proved that this numerical instability occurs occasionally in a few regions of the simulation domain situated very near to NS surface. Therefore, we have added a calculation rule for particle movement: if, during a single time step, the Lorentz factor is multiplied by a given factor (of the order of 5, but this can be adjusted!), the instability is occurring. Then, it is considered that the inertia of the particle is of no importance (in spite of being the cause of the numerical instability), an asymptotic Lorentz factor $\gamma_a$ is established to ensure equality between the electrical acceleration force and the radiation reaction force:
\begin{eqnarray} \label{eq_gamma_saturation}
 \gamma_a= \left(\frac{3}{2}  \frac{E_\parallel}{q C_m^2}\right)^{1/4}.
\end{eqnarray} 
The particle Lorentz factor is updated with this value, and the direction of motion is conserved.  At the next iteration of the particle trajectory, the computation based on Eq. \ref{eq_mouvement_lent_fondamentale} and on $\vec{E}_R$ is applied again, without necessarily resorting to Eq. (\ref{eq_gamma_saturation}). 

In the present case, Eq. \ref{eq_gamma_saturation} provides the particle energy in particular cases where the use of Eq. \ref{eq_power_curvature} would conduct to a numerical instability. Actually, the systematic use of Eq. \ref{eq_gamma_saturation} for the computation of the particle energy (that is not implemented here) would constitute a reasonable approximation, and was used in \cite{Higgins_1997}.

\subsection{Particle transport and charge deposition} \label{sec_particle_transport}

The transport of charge and current densities by particles is treated in a specific way, taking into account the stationary nature of the solution sought, the fact that an iteration does not correspond to a specific duration, the fact that each time step in a particle trajectory may be different from the others, and that the volume of each cell in the simulation domain may be different from the others. A certain formalism is needed to describe the algorithm that takes all these facts into account.

Let us consider a set of particles that have been created at time $t_0$, or have emerged from a boundary, and that have propagated since then in the absence of creation, or destruction or scattering between particles. The conservation of the distribution function along their trajectories implies that for any time $t_i$,
\begin{eqnarray} \label{eq_transport_f_pratique_details} 
& &f(r(t_i), \theta(t_i), \phi(t_i), U(t_i), s(t_i), t_i) \\
&  &=  f(r(t_0), \theta(t_0), \phi(t_0), U(t_0), s(t_0), t_0).\nonumber
\end{eqnarray} 
For the sake of economy,  we characterize the position in the phase-space with a vector $x= (r, \theta, \phi)$, and $u=(U, s)$, $x_0 = x(t_0)$ and $x_i = x(t_i)$.
Equation \ref{eq_transport_f_pratique_details} writes simply
\begin{equation} \label{eq_transport_f_pratique_x}
f(x_0,u_0,t_0) = f(x(t_i),u(t_i),t_i),
\end{equation}
where  $x(t_i)$ is the trajectory of particles initially at $x_j$ and $u(t_i)$ is its momentum in the direction along the local magnetic field.
The particle trajectory is computed with the solver mentioned in section \ref{sec_motion}, and produces the transformation $(x_0,u_0) \rightarrow (x_i,u_i)$ for any time $t_i > t_0$.
We consider that $f$ is known at time $t_0$ at every node $j$ of the phase-space grid. 

The number of particles $\delta N = f \delta V \delta U$ is constant along the trajectory, where $\delta V$ is the volume in space and $ \delta V \delta U$ is the phase-space volume. Consequently (Liouville theorem), the element of phase-space volume $\delta V \delta U$ volume is conserved. 
We note $\delta V(t_i) \delta U(t_i)$ and abbreviate $\delta (UV)_{i}$ the element of volume associated with these particles at time $t_i$, and $\Delta (UV)_{i}$ the element of volume associated with the phase space grid cell $\Omega(t_i)$ used in the code, that includes the position $x_i$ of the particles at time $t_i$. 

Let us consider the cell $\Omega_0=\Omega(x_{0})$. 
The particles initially in that cell have moved into another cell $\Omega(x_{i})$.  
The conservation of the particle number created at time $t_0$ in the phase-space cell $\Omega_0$  writes $\Delta N = f(x_0,t_0) \Delta(UV)_0= f(x_i,u_i,t_i) \delta (UV)_i$. At time $t_0$ we consider that the volume $\delta (UV)_0$ physically occupied by the particles is those of the cell space $\Delta (UV)_0$.
The volume $\delta UV(x_i,u_i,t_i)$ physically occupied by the particles is not necessarily equal to the phase-space volume $\Delta (UV)_{i}$ of the grid cell corresponding to the position of these particles at time $t_i$. A set of particles $p$ created at time $t_0$ in the cell $\Omega_0$ then contribute to the particle number in phase-space cell $\Omega(x_{t_i})$ to an amount
\begin{equation}\label{eq_DeltaN}
\Delta N(x_i,u_i,t_i) = f_p(x_0,u_0,t_0) \Delta(UV)_0,
\end{equation} 
and to the particle phase-space density
\begin{equation}\label{eq_f_contribution}
 f_p(x_i,u_i,t_i)=\frac{\Delta N(x_i,u_i,t_i)}{\Delta (UV)_{i}} = \frac{f(x_0,u_0,t_0) \Delta(UV)_0}{\Delta (UV)_{i}}.
\end{equation}
Whatever of these two quantities $\Delta N$ or $f$ can be added into an array that has the same dimensions as the phase-space grid. 
The total phase-space density is the sum of the contribution of all the source terms $f(x_0,u_0,t_0)$ associated with the phase-space grids $\Omega_0$ for pseudo particles arriving in the phase-space grid $\Omega_j$
\begin{equation}\label{eq_f_total}
 f(\Omega_i) = \Sigma_{\mathrm{ All } \,(x_0, u_0) /   (x_i,u_i) \in \Omega_j} \frac{f(x_0,u_0,t_0) \Delta(UV)_0}{\Delta (UV)_{i}} .
\end{equation}
Actually, we place not one but $N_\mathrm{part}$ particles initially in each phase-space cell $\Omega_0$. Their positions into that cell are computed randomly with a uniform law. This is done in order to reduce the particle noise that correspond to a finite sampling of the initial positions of the particles. The sum in Eq. \ref{eq_f_total} includes these $N_\mathrm{part}$ particles in each cell $\Omega_0$.

The charge density is $\rho = \int q_s f(x_i,u_i,t_i) d u_i$.
The contribution of a trajectory to the total charge  in the spatial grid element $X_i$ corresponding to $\Omega_i$ is
\begin{equation}
 Q(X_i,t_i) = \Sigma_{\mathrm{ All } \,(x_0, u_0) / (x_i) \in X_i} q_s {f(x_0,u_0,t_0) \Delta(UV)_0}.
\end{equation}

Remark : in eq. \ref{rho_x_t}, and subsequent formulas, $(x_i,u_i) \in \Omega$ is replaced by $(x_i) \in X_i$.

The contribution to the charge density is 
\begin{equation} \label{rho_x_t}
 \rho(X_i,t_i) = \Sigma_{\mathrm{ All } \,(x_0, u_0) /(x_i) \in X_i}  \frac{q_s f(x_0,u_0,t_0) \Delta(UV)_0}{\Delta V(t_i)},
\end{equation}
and the same normalization coefficients apply to the current density, and for diagnostics, to the particle density and the mean velocity (to be normalized in fine by the particle density). 

\subsection{Three arrays that represent the particle phase space flux}
In the code, the connexion between particles and charge and current densities is based on a 4D array called {\tt FFF} that contains the flux of particles, 
\begin{equation}
 {\tt FFF}(\Omega_i)=\Sigma_{\mathrm{ All } \,(x_0, u_0) / (x_i,u_i) \in \Omega_i} \frac{f(x_0,u_0,t_0) \Delta(UV)_0}{\Delta t_0}.
 \end{equation} 
This corresponds to the sources $\Delta N_0$ divided by the time $\Delta t_0$ necessary to cross the cell at the start of iteration {\tt istep}. It does not contains the effect of particle transport, therefore, this is not yet the value of $\Delta N(x_i,u_i,t_i)$.
The contribution of the trajectories to the electric charge is the flux multiplied by the time step of the trajectory element $t_i$,
\begin{equation}
 Q(X_i,t_i)=\Sigma_{\mathrm{ All } \,(x_0, u_0) / (x_i) \in X_i} {\tt FFF}(x_i,u_i,t_i) q_s \Delta t^i,
 \end{equation}
 and the contribution to charge density is
\begin{equation}
 \rho(X_i,t_i)=\Sigma_{\mathrm{ All } \,(x_0, u_0) / (x_i) \in X_i} {\tt FFF}(x_i,u_i,t_i) \frac{q_s \Delta t^i}{\Delta V(t_i)}.
 \end{equation} 
Because the model is stationary, we need the average values of the total charge and of the charge density: 
\begin{equation}
 Q(X_i)=\frac{1}{T}\Sigma_{\mathrm{ All } \,(x_0, u_0) / (x_i) \in X_i} {\tt FFF}(x_i,u_i,t_i) q_s \Delta t_i^{2},
 \end{equation}
 and the contribution to charge density is
\begin{equation}
 \rho(X_i)=\frac{1}{T}\Sigma_{\mathrm{ All } \, (x_0,u_0) / (x_i) \in X_i} {\tt FFF}(x_i,u_i,t_i) \frac{q_s \Delta t^{i 2}}{\Delta V(t_i)},
 \end{equation}
 where $T=\sum_i \Delta t_i$ is the total duration of the trajectory.
 
Another array, called {\tt FFA}, contains the flux of $\Delta N(x_i,u_i,t_i)$ at the end of the iteration {\tt istep}. Up to now, {\tt FFA} is computed to test the feasibility of the computation, but not yet used. The array {\tt FFA} will be used when we include radiative and pair creation processes. 

 During the iteration, new source terms are computed and assigned to {\tt FFB}. For a particle trajectory ending into the simulation domain, the last position of this particle contributes to {\tt FFB}. More details are given in section \ref{sec_unfinished_trajectories}. When new electron-positron pairs are created (in future studies), they will also contribute to {\tt FFB}. 
 
We made tests of radial propagation at the speed of light, with different types of injection at the inner boundary : uniform, random, and impulsive. In all cases, the simulation box is initially empty, therefore, there must be the propagation of a front of particles. With a non-monotonic Hermite interpolation of third order, 
 the results where not very good : there was a lot of diffusion, a precursor that goes much faster than light, and a non-uniform repartition of the particles, even when they were injected at a constant rate.
It was therefore decided to use instead a rough but simple and robust method of interpolation of zeroth order, where all the particle weigh (mass, charge) is assigned to a single phase-space cell. Testing with injection at a constant rate enabled uniform particle distribution throughout the simulation domain. The major advantage of this crude interpolation method is the absence of supraluminous fronts and the speed of calculation.

\subsection{Outer boundary conditions} \label{sec_outer_bc}
A particle crossing the outer boundary of the simulation domain is not replaced, but its electric charge contributes to the total charge $Q_{tot}$ associated with the whole magnetosphere. Consequently, its value is subtracted to the total charge of the neutron star, since it has left it with no return. 

\subsection{Inner boundary conditions at level 0 for the electromagnetic field} \label{sec_inner_bc0_electromagnetic}

The inner boundary conditions are based on the following assumptions. 
The magnetic field inside the star is a centered dipole, with a possible inclination of the magnetic axis relatively to the rotation axis. The magnetic field is frozen in the inner stellar plasma (said otherwise, it turns as a rigid structure with the same rotation rate as the star). 
The NS surface is a crust composed of a lattice of ions in solid rotation with the star, and of free electrons and free ions that have the same motion as the lattice. The number of free particles (electrons and ions) is much larger than the local Goldreich-Julian particle density. Their thermal motion is negligible (including it, as we tried, does not cause significant effects).
The energy required for extracting a free particle from the surface is negligible. The surface of the star is in equilibrium with the inner part of the star and with the magnetosphere. Above the surface, there is a continuous but possibly brutal particle acceleration along the direction of the magnetic field. At the surface, the particles are simply co-rotating with the star.  The transition from the corotation velocity at the surface and the plasma velocity immediately above it is continuous. This means that the perpendicular velocity and the associated electric current, vary continuously. 
 The system constituted of the neutron star and its magnetosphere possesses a null total electric charge.

Practically, we consider the internal boundary conditions (BC) at three levels.  The level inside the star corresponds to the superscript $<$, and it not included in the numerical simulation grid. 
Level 0 is the top of the neutron star crust and corresponds physically to superscript $>$, corresponding to an altitude $z=0$, a distance $R_*=r_0$ from the center of the neutron star, and a radial  grid index equal to 0. 

For practical reasons concerning the acceleration of particles detailed in the next section, we also consider the level right above it, of index 1. 
The interval between these two levels 0 and 1 contains the lowest part of the pulsar atmosphere. As mentioned in section \ref{sec_grid}, the difference of altitude ${\rm d} r=r(1)-r(0)$ between these two levels is of the order of one centimeter. 

Let us first concentrate on the electromagnetic field at level 0, that constitutes the boundary on which the assumptions opening this section are applied. 

Since the equations of electromagnetism are linear, we can consider separately the phenomenon that intervene in the electric charge and current densities, and add their respective contributions to the boundary conditions.
The partition of electromagnetic sources used in the present work is the same as in \cite{Petri_2002} and is the addition of the effect of the rotation of the magnetized sphere, of the net electric charge inside the star, and  of the charge and current densities outside of the star. 
Therefore, the electric field at any point of the NS surface, of spherical  coordinates $r_0,\theta,\phi$ for any $\theta$ and $\phi$, is 
\begin{equation}
\vec{E}^> = \vec{E}_{v}^>+ \vec{E}_{Q_t}^>+ \vec{E}_{\rho}^>
\end{equation}
where the exponent $>$ is relative to any position at the NS surface, $\vec{E}_{v}$ is the vacuum solution for a rotating magnetized star, $\vec{E}_{Q_t}$ is caused by the electric charges inside the star, and $\vec{E}_{\rho}$ is caused by the charge density outside of the NS and of its surface. There is a similar decomposition for the magnetic field. 

The most specific effect regarding neutron stars electrodynamics is those of rotation and magnetic field. Because it can be treated independently form those of the charge and current densities outside of the star, we consider the rotating conducting magnetized sphere into an empty volume. The internal magnetic field fits the dipole solution, and the internal electric field is the corotation electric field $\vec{E}= \Omega R_* \sin \theta \; \vec{e_\phi} \times \vec{B}$. With explicit components of the corotation electric field, and the continuity conditions of electromagnetism     
 \begin{equation}
 B_{r,v}^{>}=B_r^{r<}, \; E_{\theta,v}^>=E_\theta^<= -(\Omega/c)R_* B_r^< \sin \theta, \; E_{\phi,v}^> =0. 
 \end{equation}
 
The other components can vary discontinuously. 
The solution to this problem for a sphere embedded into a vacuum environment is known analytically, and we use the relations given in \citep{Mottez_2015c} to define the values of $E_{r ,v}^{>}, B_{\theta ,v}^{>}$ and $B_{\phi ,v}^{>}$. These equation are provided for magnetic multipoles, and for a magnetic dipole. In the simulations presented in sections \ref{aligned}-\ref{oblique}, this corresponds to the dipole solution given in \citep{Deutsch_1955}. Let us notice that this solution implies a discontinuity of the electromagnetic field,  charge and current surface-densities that are not at all negligible. Nevertheless, in practice,  we do not need to evaluate the  4-charge surface-density.  

Possibly, the overall electric charge contained in the magnetosphere is finite, we note $Q_{t}$ its value. The charge contained in the NS star is therefore $-Q_t$. Because the NS is an equilibrium conducting volume, the electric field is null (we have removed the effect of corotation) and the electric charge $-Q_t$ lies at the surface of the star. Is this surface charge density uniform ? If it was not uniform, it would create tangential electric fields (relatively to the surface) that would combine with the magnetic field to produce a plasma motion. Since the corotation charge density is considered separately, the surface charge density does not induce any motion. Therefore, the surface charge density associated with $-Q_t$ is uniform, and it is the cause of a radial electric field with a spherical symmetry, and equal to $E_{r,Q_t}^>=-Q_t \times r^{-2}$. The two other components $E_{\theta,Q_t}^>$ and $E_{\phi,Q_t}^>$ are null. 

The last constraint on the electromagnetic field at the NS surface is given by a 4-charge density distribution outside the neutron star. In the present paper, we adopt two different hypothesis to deal with that charge density distribution. 

In sections \ref{aligned}, we simply consider that this charge density is weak and induces a negligible electric field $\vec{E}_\rho^>$ at the NS surface. This hypothesis is met in \cite{Higgins_1997}. 

In section \ref{aligned_electrosphere_full}, we do not assume the weakness of the charge density, but we consider that the associated electric field is, in a first approximation, of electrostatic nature, as in \cite{Petri_2002}. Then, a first solution $\vec{E}_{\rho}=-\nabla \Phi$ is computed in the whole simulation domain, under the electrostatic hypothesis.
The electric potential is the solution of $\nabla^2 \Phi= -4 \pi \rho$ where $\rho$ is the electric charge density contained in the simulation domain. This equation is solved with the same spectral method as for the other equations of the electromagnetic field. The part inside the neutron star (it is not contained in the simulation domain) is a conducting area, its electric potential is uniform. Since the charge $-Q_t$ inside the NS is considered separately, the electric potential $\Phi$ is the same at the NS surface and at infinity. We fix its value to zero: for any $\theta$ and $\phi$ coordinates, $\Phi(r_0,\theta,\phi)=\Phi(r_{max},\theta,\phi)=0$.  As a consequence, at the NS star surface, only the radial component ${E}_{r,\rho}^>$ of $\vec{E}_{\rho}^>$ is finite, the two other components are null. In a second stage, once ${E}_{r,\rho}^>$ is computed, it is set as the contribution of the magnetospheric charge density $\rho$ to the boundary condition at the NS surface (we don't "use" $\vec{E}_{\rho}$ for $r \ne r_0$), and an electromagnetic solution is computed over the whole simulation domain. 

In section \ref{oblique} both boundary condition with and without ${E}_{r,\rho}^>$ are used, and provide equal electrosphere solutions. 

\subsection{Inner boundary conditions at level 0 for the particles} \label{sec_inner_bc0_particules}

The injection of particles from the NS surface requires the determination of their energy (or their velocity) and their density. 

At level 0, the particles constituting the plasma in the crust have a negligible kinetic energy, and reasonably, in the corotating reference frame, we can set $\gamma=1$ at level 0. Giving them random energies corresponding to the thermal dispersion (i.e. adding fractions of unity to $\gamma$) in the crust does not change significantly their Lorentz factor relatively to the values that are reached at high altitudes with or without this correction. 

The particles initially at the surface of the star are accelerated by the electromagnetic field. We assume that the number of particles with electric charges of both signs is very large on the NS surface. Because of the NS rotation, the difference between the density of positively charged particles and negatively charged particles results in the Goldreich-Julian charge density. These particles are exposed to the surface electric field. This electric field, when of finite modulus, pushes above the surface only particles having one the two signs. The driver of this electric field is the NS rotation, and for this reason, and because of the continuity of the plasma corotation with the star surface, an infinitesimal acceleration must allow the injection of the GJ charge density  at an infinitesimal distance from the surface. Therefore, the density of injected particles is the GJ density. Of course there are other sources of particles : particles created in the magnetospheric plasma (pair creation in pulsar magnetospheres), particles coming from interstellar space or from a neighboring star (non isolated NS), and particles previously injected from the surface into the magnetosphere that come back into the NS.

Ideally, these conditions are sufficient to define the inner boundary conditions. We could compute the particle acceleration between levels 0 and 1 using the algorithm introduced in sec. \ref{sec_motion}. Unfortunately, as was experienced in early versions of the code, because of the huge parallel electric field existing right above the neutron star's crust, this require high numbers of time steps that  cost a lot of CPU time. To overcome this problem, we compute analytically the particle acceleration over a distance ${\rm d} r=r_{1}-r_0$ (also noted $d$) that is the radial extension of the first grid cell. This is why level 1 is the object of a specific treatment.  

\subsection{Inner boundary conditions at level 1} \label{sec_inner_bc1}

In the transition from level 0 to level 1, that correspond to an altitude jump of the order of one centimeter, the $B_r, E_\theta, E_\Phi$ components of the electromagnetic field do not change significantly because they do not depend on the surface charge and current. We set them with the same value at levels 0 and 1. 

The  components $B_\theta, B_\phi$ and $E_r$ at level 1 depend on the charge charge and the electric current circulating in the atmospheric plasma between level 0 and level 1. We cannot assume any continuity for them. The electromagnetic field components $B_\theta, B_\phi$ and $E_r$ at level 1 are set with their values computed during the previous iteration. At first iteration, $B_\theta, B_\phi$ and $E_r$  correspond to the vacuum solution, that for a dipole, can found in \cite{Deutsch_1955}.

The motion of particles between levels 0 and 1 is computed analytically. By the mean of Lorentz transformations, we can compute the local electromagnetic field in the frame rotating with the star. Values in this frame are denoted with a prime.   
The radiation reaction force is neglected.
After electric acceleration over at time ${\rm d}t$, the energy is $\gamma_0= q E_{//}'  d/mc^2 +1$, where  $E_{//}'$ is the parallel electric field in the co-rotating frame, right above the surface and $d=r(1)-r(0)$. We neglect the magnetic field inclination, and the parallel particle velocity  in the co-rotating frame is
\begin{equation}\label{eq_v_sur_c}
\frac{v_{0}'}{c}=\frac{\sqrt{(q E_{//}' d+mc^2)^2 -(mc^2)^2}}{(q E_{//} 'd+mc^2)}.
\end{equation}
When ${v_{0}'}$ is known, the  4-velocity is expressed in the observer's frame where all the other computations are performed. 
At the surface, $n_\mathrm{surface}=n_\mathrm{GJ}$ is the Goldreich-Julian density of particles in a conducting ideal plasma in rotation at the pulsar spin rate. 
Then, we have, at level 1 above the surface, a distribution of particles with a non negligible amount of energy. We use this distribution as the injection condition from the star crust into the magnetosphere. From this surface containing accelerated particles, we can "launch" the particles, and compute their trajectories.
The subsequent values of the 4-velocity are computed using the numerical method based more traditionally on Taylor expansion (sec. \ref{sec_motion}). 
Then, the subsequent time-step $\Delta t_1$ is forced to fit the usual Courant condition $\Delta t_1 < d/v_0 \ll \Delta t_0$.

This simplified computation has an imperfection: the vertical distance $d$ was used in Eq. (\ref{eq_v_sur_c}) while the local inclination of the magnetic field (its $\theta$ and $\phi$ components) could have been considered. This approximation minimizes the energy of the particles injected at level 1.  Anyway, we checked the values of the particles Lorentz factors injected at level 1. They are typically comprised between 1 and a few $10^4$. These values are negligible compared to the Lorentz factors taken subsequently by the particles, with typical values above $10^6$. Therefore, this imperfection in the particle injection process do not change significantly the physics of our simulated electrospheres .    


Let us consider first that one quasi-particle is injected for each cell. This particle represents the whole physical particle population in this cell.
At the star's surface, inner boundary condition corresponds to particle injection over the predetermined distance ${\rm d}r$ corresponding to the difference of altitude between levels 0 and 1.
\begin{equation}\label{eq_inner_BC_particle_injection}
{\rm d}f_\mathrm{3, source}= f(x_0,t_0)=  {n_\mathrm{surface} S {\rm d}r},
\end{equation} 
where $S$ is the surface element $S=r^2 \sin(\theta)d\theta d\phi$,  $n_\mathrm{surface}$ is the particle number density. In practice, the surface element is $S=R^2\, \Delta( - \cos(\theta)) \, \Delta\phi$. The index $3$ of ${\rm d}f_\mathrm{3, source}$ means that this quantity that depend only of the 3D position is integrated over momentum (it is not yet defined in the 4D phase space).

Then $f$ corresponds to the number of particles in each phase-space cell that is injected during the time step ${\rm d} t_0={\rm d}r/v_0$ where  $v_0$ is the particle velocity. 

This distribution ${\rm d}f_\mathrm{3, source}$ is integrated over the energies. We need a distribution that takes the energy dependency into account. In the simulations presented in this paper, if one particle energy represents the whole local population, the distribution is simply null for any value of $u$, except $u_0$, 
\begin{equation}\label{eq_inner_BC_particle_injection_1part}
{\rm d}f_\mathrm{source}= f(u_0, x_0,t_0) =  \delta(u-u_0){\rm d}u \; {n_\mathrm{GJ} S v_0} {\rm d}t,
\end{equation}
where $\delta(u-u_0)$ is the Dirac distribution centered on $u_0$. Let us note that $\delta(u-u_0){\rm d}u$ is a dimensionless number. Practically, the momentum space is discretized (sec. \ref{sec_grid}) and $\delta(u-u_0)=1$ when $u$ is in the bin containing $u_0$ and $\delta(u-u_0)=0$ otherwise. 

Actually, we do not consider only one particle per inner boundary cell. We start with a bunch of $N_\mathrm{part/cell}$ particles per cell, each of them having the same statistical weight, with slightly similar initial radial distances $d_i$ in the interval $[0,d]$. 
Equation \ref{eq_inner_BC_particle_injection_1part} is therefore replaced by 
\begin{equation}\label{eq_inner_BC_particle_injection_2}
{\rm d}f_\mathrm{source}= (N_\mathrm{part/cell})^{-1} \Sigma_{i \in [1,N_\mathrm{part/cell}]}   \delta(u-u_i){\rm d}u  \; {n_\mathrm{GJ} S v_i} {\rm d}t,
\end{equation}
and $u= v_{i,\parallel} \gamma_i$ and $v_i$ is computed in the same way as $v_0$. (This could be refined.) The initial values of the particle co-latitudes and azimuths are also random, with values taken in the volume defined by $\Omega_0$.

\subsection{Injection of particles corresponding to unfinished trajectories} \label{sec_unfinished_trajectories}
A particle trajectory can leave the simulation domain during an iteration. It is then interrupted. It is also possible that a trajectory reaches the maximum number $P_{max}$ of allowed elements and is still in the simulation domain. In that case, the trajectory is restarted at the next iteration. The re-injection occurs at the location and with the parallel momentum of the last element, and the value assigned to the flux $\tt FFB$ of particles to be re-injected is
\begin{equation}
{\tt FFB}(x_i,u_i,t_i)=f(x_0,u_0,t_0) \frac{\Delta t_0}{\Delta t_i},
\end{equation}only
where $i=P_{max}$.
At the end of an iteration, $\tt FFB$ is copied into $\tt FFF$ that contains both the sources  on the inner boundary, and the sources located inside the simulation domain associated with non null values of $\tt FFB$. Then, $\tt FFB$ is set to zero for the next iteration.

%

\section{Aligned electrospheres in the low density approximation} \label{aligned}

In this section, the electric field induced by the electrosphere charge density $\vec{E}_{\rho}^>$ is assumed to be much less than $\vec{E}_{v}^>$ and $\vec{E}_{Q_t}^>$, and is neglected. 
   \begin{figure}
   \centering

   \includegraphics[width=8cm]{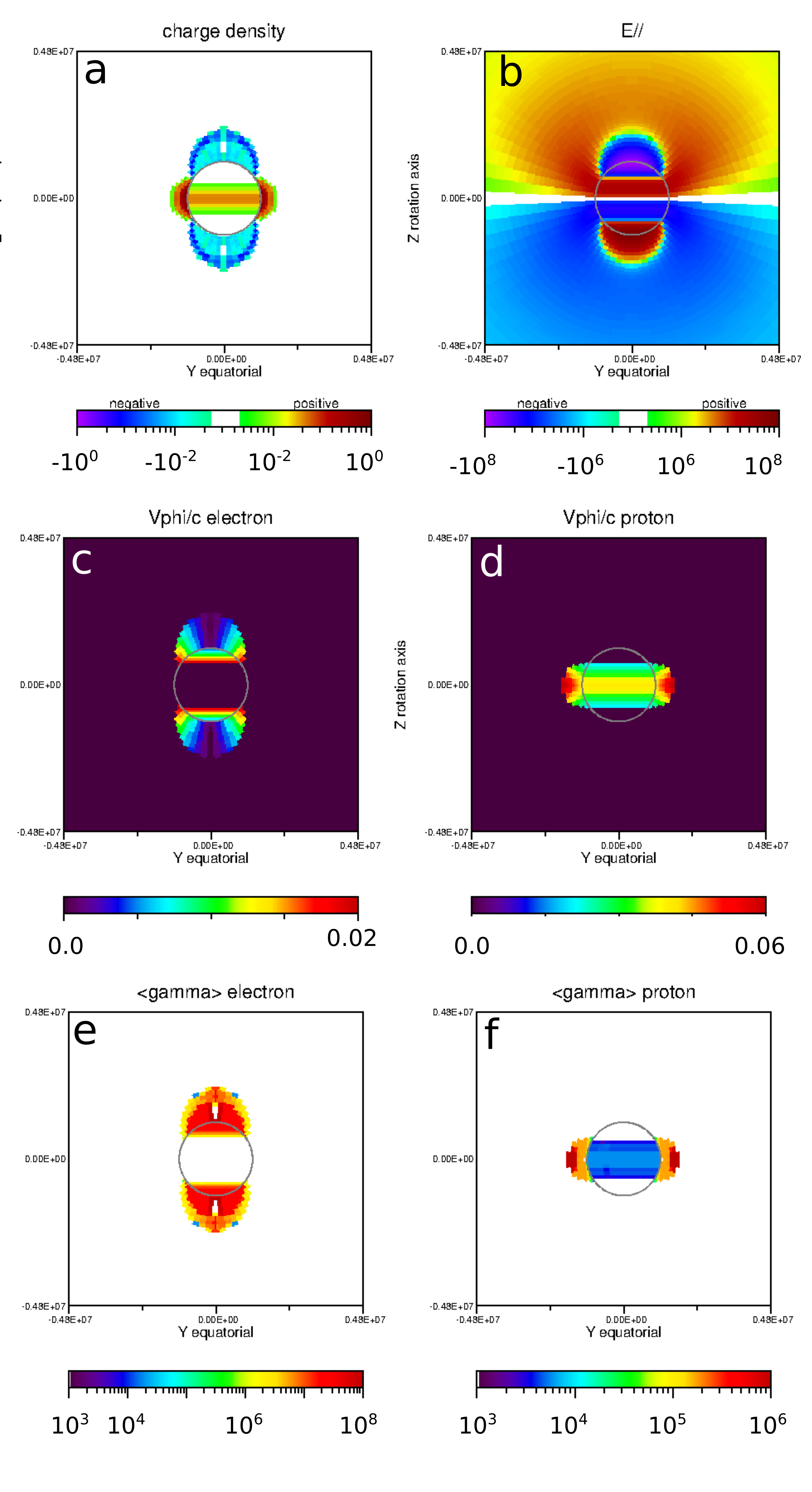}
   \caption{Aligned electrosphere with surface magnetic field $B_*=10^{9}$G, rotation period $P=10$ms, and a surface composed of electrons and protons. The inner circle represents the neutron star radius $R_*=12$km. Data cross-section in the plane including the rotational axis $Oz$ and the magnetic axis of symmetry, except inside the inner circle where it represents data at 20 meters above the surface.  (a) charge density including the contribution of ions and electrons (b) parallel electric field (c) electron azimuthal velocity (normalized to the speed of light $c$) (d) proton azimuthal velocity (e) mean electron Lorentz factor (f) mean proton Lorentz factor.  }
              \label{fig_aligne_I00_B1E9G_P10ms}%
    \end{figure}

The first full capacity test of \textit{Pulsar ARoMa} consists of the simulation of aligned electrospheres, and their comparison with other aligned electrospheres in the literature. A series of simulations was carried out. The one shown in Fig. \ref{fig_aligne_I00_B1E9G_P10ms} corresponds to a star's radius $R_*=12$ km, a surface magnetic field $B_*=10^9$ G, and a period $P_*= 10$ milliseconds. All the other parallel electrospheres simulated with \textit{Pulsar ARoMa} present similar features.

 We can see with the charge density map (figure a) that the electrons form two domes, one above each pole. The protons form a small equatorial belt, or a torus. This structure with two polar domes and an equatorial belt is characteristic of all the aligned electrospheres described in the above cited literature.

 The parallel electric field is plotted on the figure (b). The parallel electric field is characterized by two domes with a given polarity, inserted into a larger region of opposite polarity. The two regions are separated by a "polarity inversion dome" (transition from red to blue on fig. \ref{fig_aligne_I00_B1E9G_P10ms}, above each polar region) where its sign reverses. This parallel field extracts the electrons at the bases of the two domes, and it extracts the protons near the equatorial plane. Then, the particles combine a motion of high velocity along the magnetic field lines with drifts guided by the corotation electric field and magnetic field line curvature. The protons are quite soon re-injected onto the NS, at the hemisphere opposite to that of their emission. The electrons are accelerated above the NS surface, and reach the limit of the electric polarity inversion dome, and then are accelerated back into the direction of the star. Then they cross again the electric polarity inversion dome, they are decelerated. In spite of the radiative losses, most of them reach the NS. This dynamic behavior is not those of a force-free plasma, but the resulting shape of the electrosphere is very alike that of those described in \citet{Petri_2002} that is based on the force-free hypothesis.

The total electric charge of the magnetosphere converged in about 2 iteration toward a value $Q_t=-0.36 \times 10^{-3} Q_c$, where $Q_c=R_*^3 \Omega_* B_*/ 3c$. The charge $Q_c$ is the same (in Gaussian units) as those in \citet{Petri_2002}. In Petri, the total charge is a model parameter, and in the present work, it is computed self-consistently.  Therefore, direct comparisons are not easy. Nevertheless, our simulation would more or less correspond to those in Petri with $Q_c = 0$. The domes in our simulation are smaller but comparable to the ones in Petri. \citet{Petri_2002} also tired values of $Q_c$ of the same order of magnitude as $Q_t$, and this is not relevant in comparison with the present simulations.  

The mean azimuthal normalized velocities $v_\phi/c$ are plotted for the electrons on figure (c) and for the protons on figure (d). We can see, as in \citet{Petri_2002} that the proton velocity reaches its maximum at a distance of the order of a few tens of percents of the NS radius. We can also see that the electron mean velocity increases with co-latitude. 

The particle energies are not documented in the literature. In plots (e) for electrons and (f) for protons, we can see that Lorentz factors reach values up to $10^{8}$. In the "polarity inversion dome", the mean energy of electrons is lower by an order of magnitude. 
This is normal, as it's at this point that the particles turn in the direction of the star, therefore, in the "polarity inversion domes", their parallel velocity temporally reaches a null value and the average energy is lower.
 
\section{Aligned electrosphere with electrostatic boundary conditions} \label{aligned_electrosphere_full}
\begin{figure}
   \centering
     \includegraphics[width=8cm]{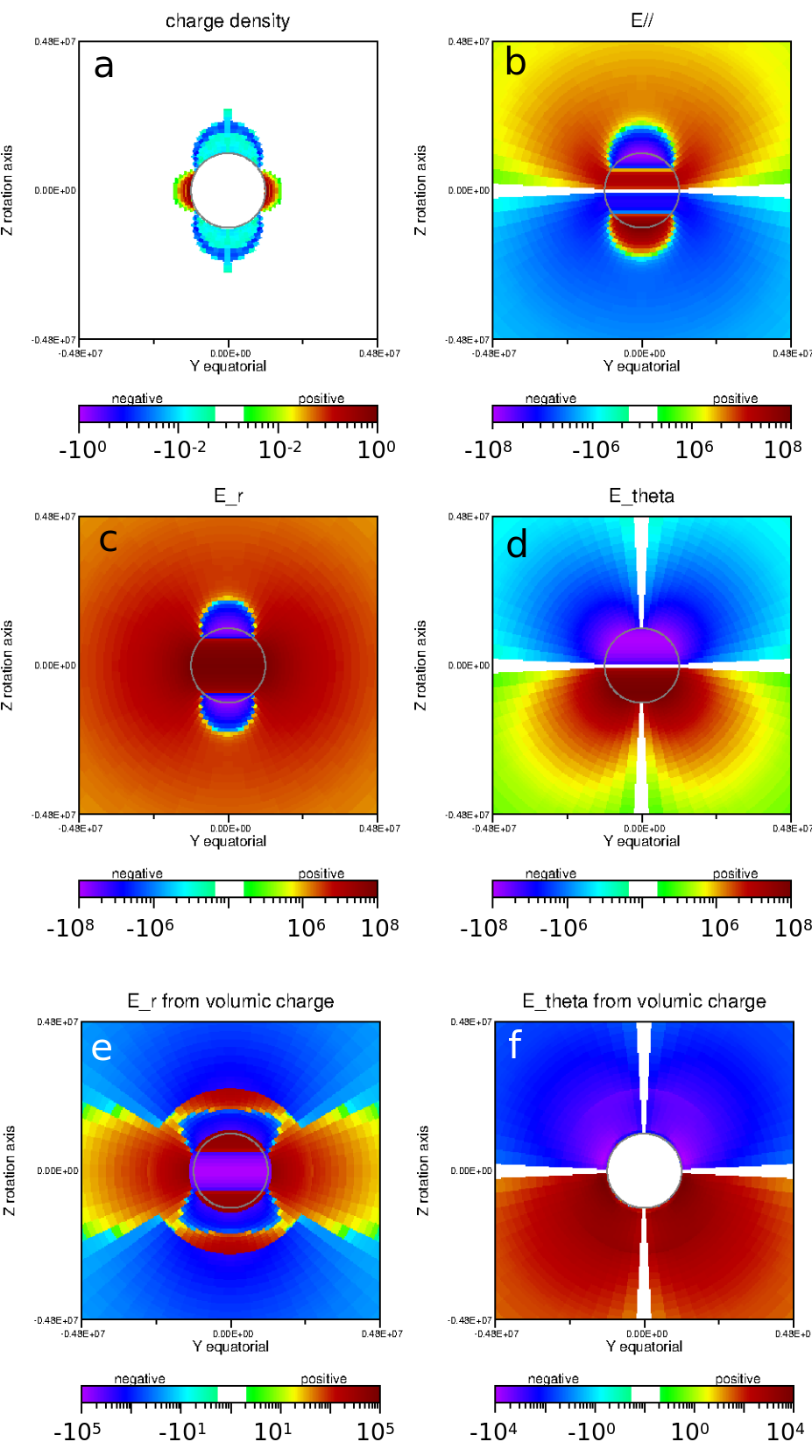}
   \caption{Aligned electrosphere with the same parameters as in Fig. \ref{fig_aligne_I00_B1E9G_P10ms} but with internal boundary conditions that take into account the charge density of the electrosphere. Cross-section of data in the plane including the axis of rotation $Oz$ and the axis of magnetic symmetry, except inside the inner circle where it represents data at the surface of the NS star.  (a) charge density including ion and electron contributions (b) parallel electric field, (c,d) radial $E_r$ and latitudinal $E_\theta$ components of the electric field, and (e,f) same components of the $\vec{E}_\rho$ electrostatic electric field induced by the electrosphere's charge density, whose value ${E}_{r,\rho}^>$ is included in the NS boundary conditions.  }
              \label{fig_aligne_comparaison}
    \end{figure}

The electric field induced by the electrosphere charge density $\vec{E}_{\rho}^>$ is now taken into account. 
A parallel magnetosphere with the same parameters as in the previous section is simulated. The stationary solution is the same, to a very high level of accuracy. As with the vacuum boundary solution, at the second iteration $Q_t=-0.36 \times 10^{-3} Q_c$. The panels (a) and (b) in both figures  \ref{fig_aligne_I00_B1E9G_P10ms} and \ref{fig_aligne_comparaison} show no difference. The reason is that the electric field induced by the charge volume density in the electrosphere (panels e and f of fig. \ref{fig_aligne_comparaison}) reaches a maximal value of $10^5$ statV/cm, when the total electric field, that is mainly the vaccum solution of a rotating magnetized sphere, reaches values 1000 times larger (panels c and d of the same figure). The conclusion is that an aligned electrosphere can set up in the low density approximation, as was done for instance by \cite{Higgins_1997}.

\section{Oblique electrospheres} \label{oblique}

All the oblique electrosphere simulations presented in this section were carried out once in the $\vec{E}_{\rho}^> =0$ approximation and once in the $\vec{E}_{\rho}^> \ne 0$ approximation. In the latter case, it systematically turns out that ${E}_{\rho}^> \sim 10^4$ statV/cm when the total electric field ${E}^> \sim 10^7$ statV/cm, and apart from the value of $\vec{E}_{\rho}^>$, the solutions given by the two simulations do not allow us to distinguish their boundary conditions. Consequently, we no longer mention which of the two boundary conditions is used in the following presentation of oblique electrospheres. 

\begin{figure}
   \centering

   \includegraphics[width=8cm]{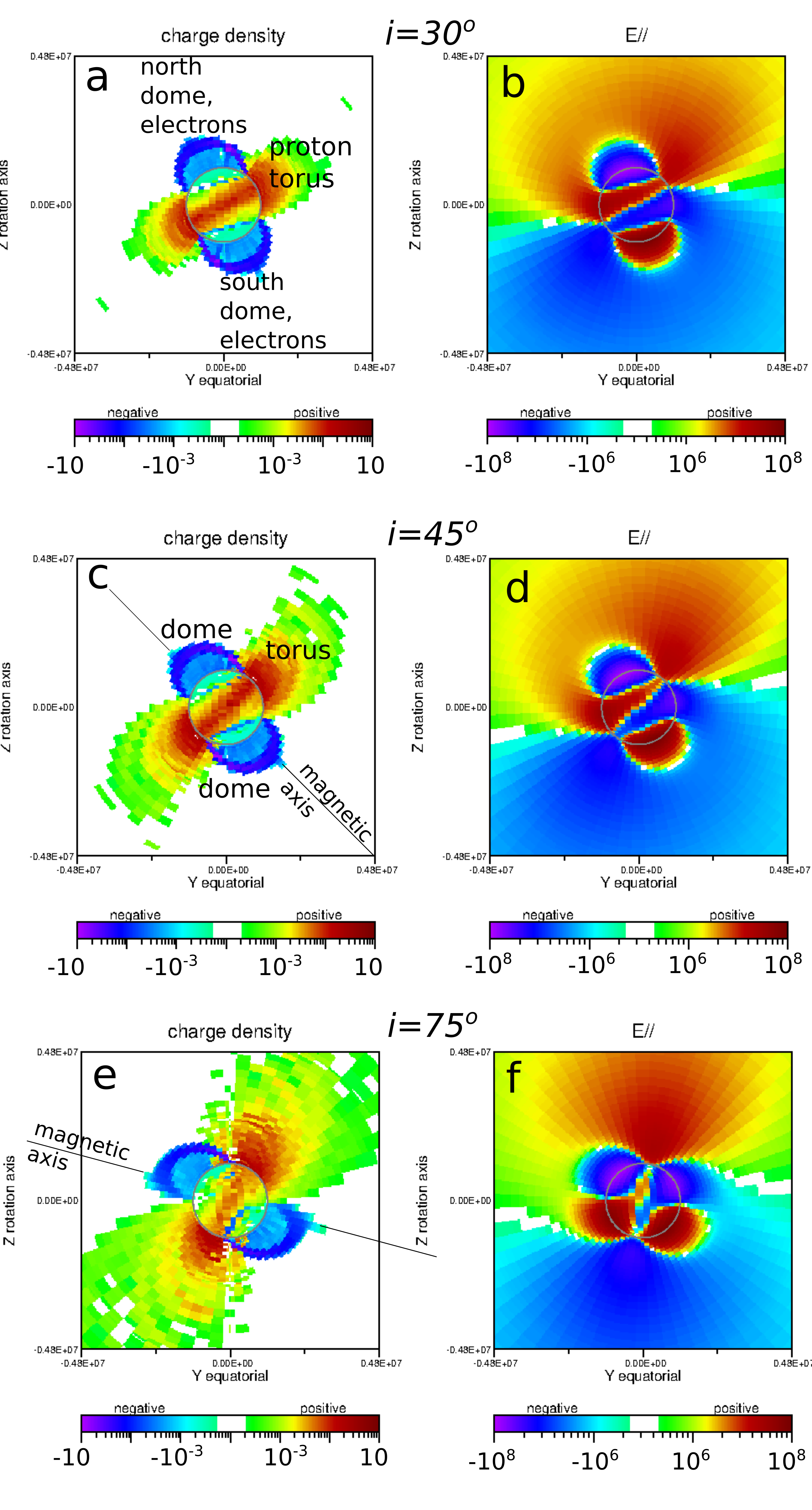}
   \caption{Charge densities and parallel electric fields for oblique electrospheres with inclination angles (a,b) $I=15^o$ (c,d)$I=45^o$ and (e,f) $I=75^o$. The other parameters and the projections are the same as in fig. \ref{fig_aligne_I00_B1E9G_P10ms}. }
              \label{fig_rho_Eparallel_30_45_75_annote}%
    \end{figure}

Figure \ref{fig_rho_Eparallel_30_45_75_annote} shows the charge densities and parallel electric field for electrospheres with the same parameters as in sec. \ref{aligned} except for their magnetic inclination angles that take the values $i=30^o$ (plots a,b), $i=45^o$ (c,d) and $i=75^o$ (e,f). If electron-positron pairs creation was effective, these parameters would correspond to a typical recycled millisecond pulsar.
Figure \ref{fig_rho_Eparallel_30_45_75_annote} shows the electric charge densities (plots a,c,e) and the parallel electric fields (plots b,d,f). 

The structure of oblique electrospheres, with two electron domes and the proton torus, are the same as with the aligned electrosphere of fig. \ref{fig_aligne_I00_B1E9G_P10ms}. The parallel electric field also has the same structure. For $i=75^o$, the parallel electric field seems more complex, with a quadrupolar structure. 
Actually, the change of sign of the parallel electric $\vec{E} \cdot \vec{B}/B$  field is caused by an inversion of the projection of the magnetic field direction $\vec{B}/B$ onto the electric field, and not by a inversion of the electric field polarity. Therefore, we can still speak of two main regions of accelerating electric field separated by an intermediate torus-like region separating the two domes that are isolated from each other. The quadrupole structure of the parallel electric field is simply an  effect of the projection of $\vec{E}$ onto $\vec{B}$. 

The particle densities of electrons and of ions have been plotted separately (not shown). These plots show electrons only in the domes, and protons only in the torus. Therefore, the negative charge densities regions that we can see in plot (a,c,d) are filled only with electrons, and the positive charge density regions are filled only with protons. 

With $i\ne 0$, the  electron domes are asymmetrical, with greater extension on the side closest to the axis of rotation, their summit lies between  the magnetic axis and the rotation axis $Oz$.  Inside the domes, the electron density is lower than in the inversion region by a decade or more. This is explained by the electron velocity that is almost $c$ when it enters the electrosphere, and that diminishes while reversing its direction in the electric "polarity inversion dome". The particle number conservation implies that in the low velocities zones, the particle number density is higher than in high velocity zones.  

The size of the proton torus is an increasing function of the magnetic inclination angle $i$. Nevertheless, we can notice that the color scale for densities covers a range of four decades, and when a range of only three decades in considered, the size of the domes and torus is restricted to about one NS radius.

\begin{figure}
   \centering

   \includegraphics[width=8cm]{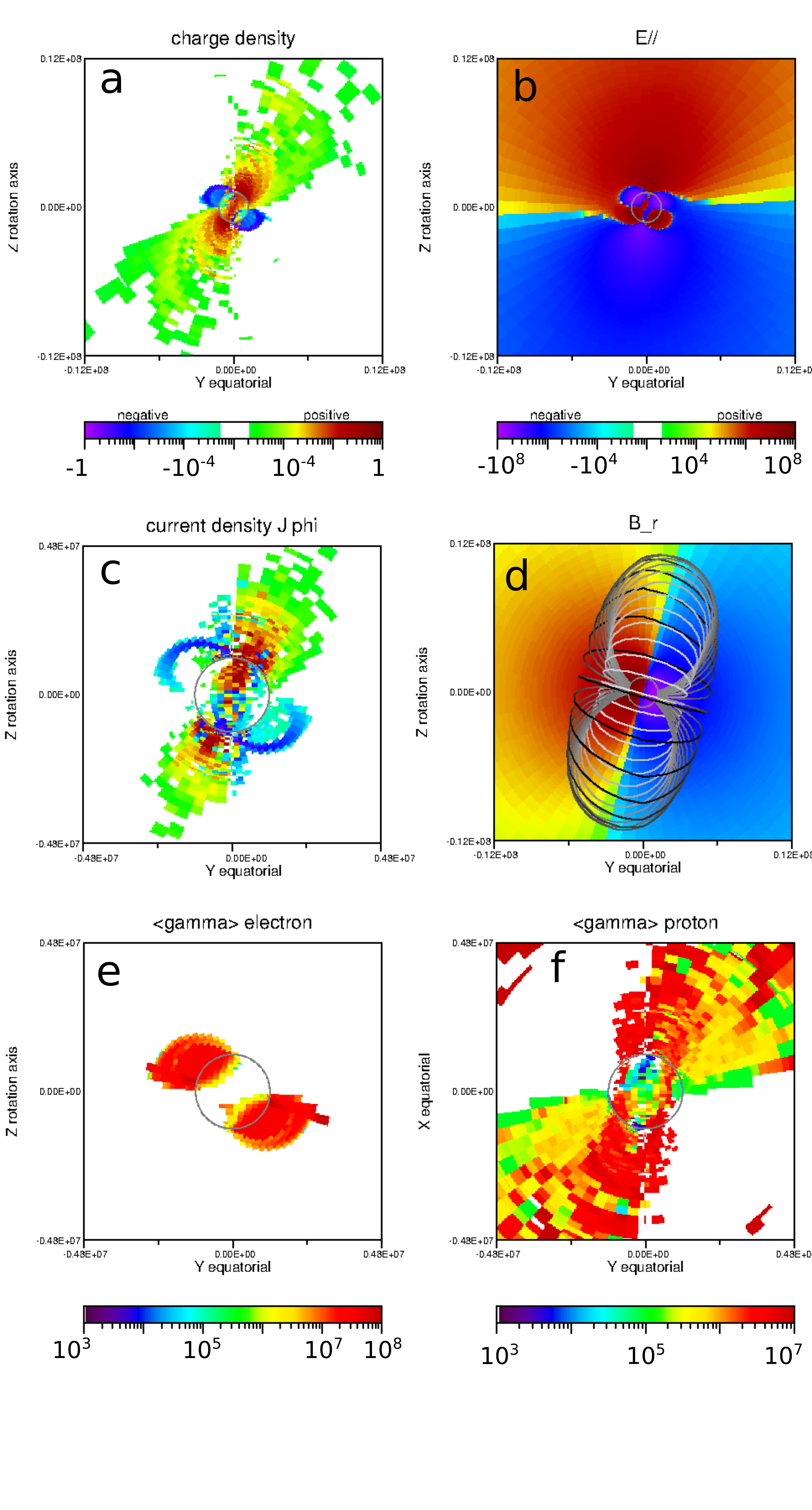}
   \caption{Oblique electrospheres with inclination angle $I=75^o$. The other parameters and the projections are the same as in fig. \ref{fig_aligne_I00_B1E9G_P10ms}. (a) electric charge density; (b) parallel electric field; (c) azimuthal current density $j_\phi$; (d) radial magnetic field and magnetic field lines; (e) mean electron Lorentz factor; and (f) mean proton Lorentz factor.}
              \label{fig_B1E9G_P10ms_I75deg}%
    \end{figure}

  More data characterizing the electrosphere with an inclination angle $i=75^o$ is presented in fig. \ref{fig_B1E9G_P10ms_I75deg}. 
  The projections angles are the same as in fig. \ref{fig_rho_Eparallel_30_45_75_annote}, but two scales are used. The scales are   
  given by the circle in the center of each plot, whose diameter equals $R_*=12$ km. Plots (a) and (b) are the  charge density and the parallel electric field, but at a larger scale. They show the large spatial extent  of the low density part of the proton belt, and the slow radial decrease of the parallel electric field. We have also checked that for distances $r > 7 R_*$, as for a central point charge, the electric field decreases proportionally to $r^{-2}$. 
  Plot (c) shows the azimuthal component of the electric current. It is highly structured, and the current density causes a weak distortion of the magnetic field relative to the vacuum solution that can be noticed by examination of magnetic field lines having their feet at the same magnetic latitude.   On plot (d) the parts of the field lines closer to the observer are in white, and the farther are in black. We can see that their shape is not symmetric relatively to the magnetic axis. It is symmetric with the vacuum solution. Nevertheless, the distortion of the magnetic field is small near the NS, and the approximation used in most models, where the magnetic field is the vacuum solution, is globally correct.
  The parallel electric field is also different from the vacuum solution. Nevertheless, the subtraction of the parallel electrosphere electric field with the central charge electric field $Q_c/r^2$ and the vacuum solution shows a structured pattern but with an amplitude that does not excess a few percents of the electrosphere electric field (not shown). 
  
  The electron and proton Lorentz factors of the oblique electrospheres are not significantly different from those of the aligned electrospheres, they reach average values of a few $10^7$ for electrons, and a few $10^6$ for protons. In terms of kinetic energies, we can notice that the protons reach larger values than the electrons. This is because the electron loose a lot of their energy through curvature radiation, while the proton radiation loss is much weaker. 
  
\begin{figure}
   \centering

   \includegraphics[width=8cm]{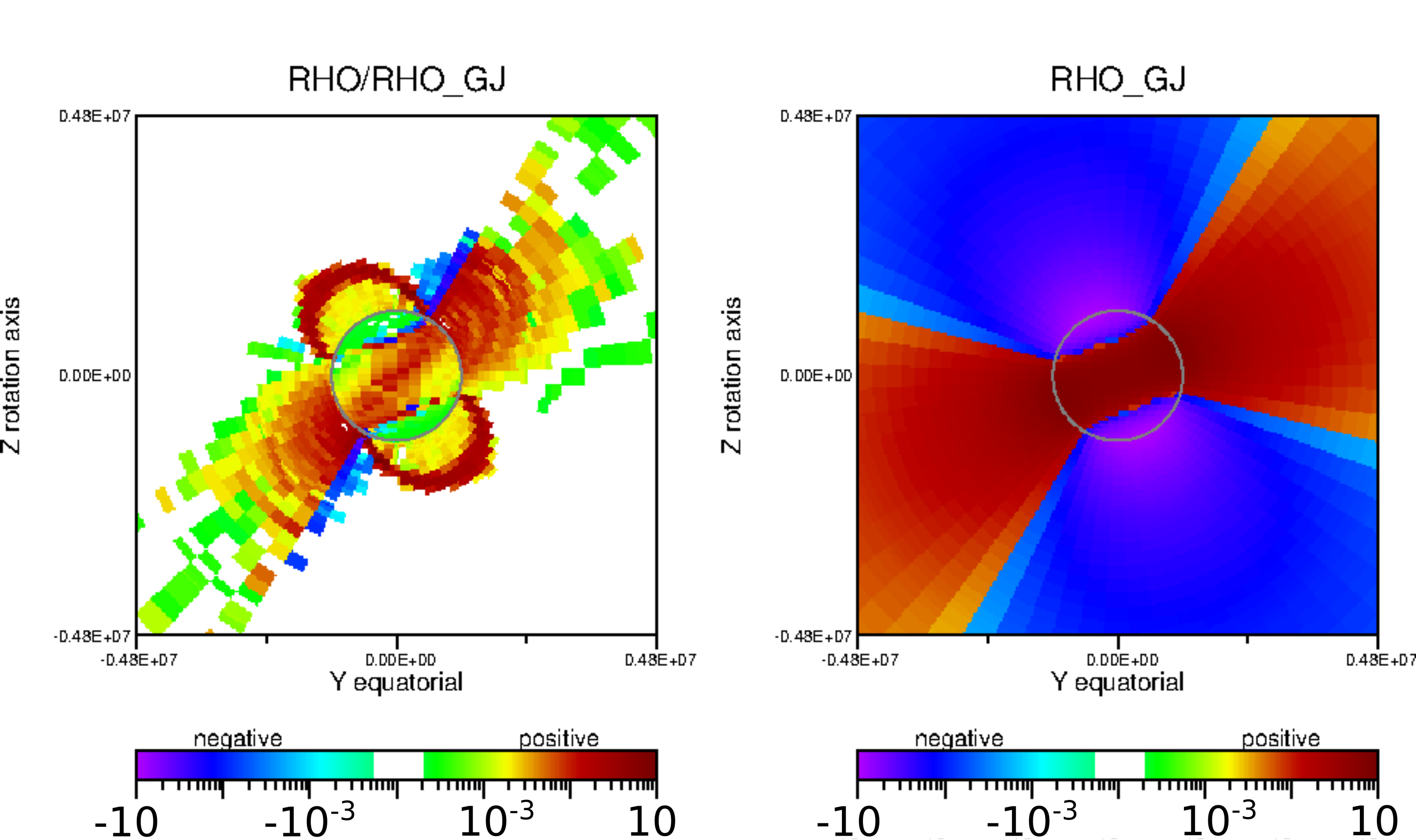}
   \caption{Oblique electrospheres with inclination angle $I=45^o$. The other parameters and the projections are the same as in fig. \ref{fig_aligne_I00_B1E9G_P10ms}. (a) Electric charge density divided by the local Goldreich-Julian density $\rho/\rho_{GJ}$, and (b) Goldreich-Julian density $\rho_{GJ}$. }
              \label{fig_Rho_sur_RhoGJ_B1E9G_P10ms_I45deg}%
    \end{figure}

As said in the introduction, the corotation plasma density, also called the Goldreich-Julian density $\rho_{GJ}$ plays an important role near the neutron star's surface. It corresponds to the inner boundary particle number density set in these simulation. So, by construction, the ratio $\rho/ \rho_{GJ}$ of the charge density over the corotation charge density is one at the neutron star's surface. This ratio is plotted in fig. \ref{fig_Rho_sur_RhoGJ_B1E9G_P10ms_I45deg}, on the left-hand-side for a pulsar with an inclination angle $i=45$ degrees (the other parameters are the same as above). On the right hand-side, the corotation charge density is plotted. Above the NS disk, plotted data corresponds to an altitude of 20 meters above the surface. We can see that the charge density in the electrosphere is comparable to the corotation density in the proton belt, and at the outer boundaries of the electron domes. Inside the electron domes, the electrons are very fast, with the same flux as at the inner boundary, therefore, the density is weak  $\rho \sim 10^{-3} \rho_{GJ}$. It might seem strange that the electrons density does not correspond $\rho_{GJ}$. Actually, it does not mean that they do not rotate with the star. When $\rho_{GJ}$ is evaluated, it is considered that the corotating frame is a frame without acceleration. In the simulations, the accelerating parallel electric field   is strong in the observer's frame as well as in the corotating field. It is therefore not inconsistent to see a ratio  $\rho / \rho_{GJ} \neq 1$ and a plasma in rotation.   

 \subsection{Influence of the radiative losses, and of the plasma composition.}
\begin{figure}
   \centering

   \includegraphics[width=8cm]{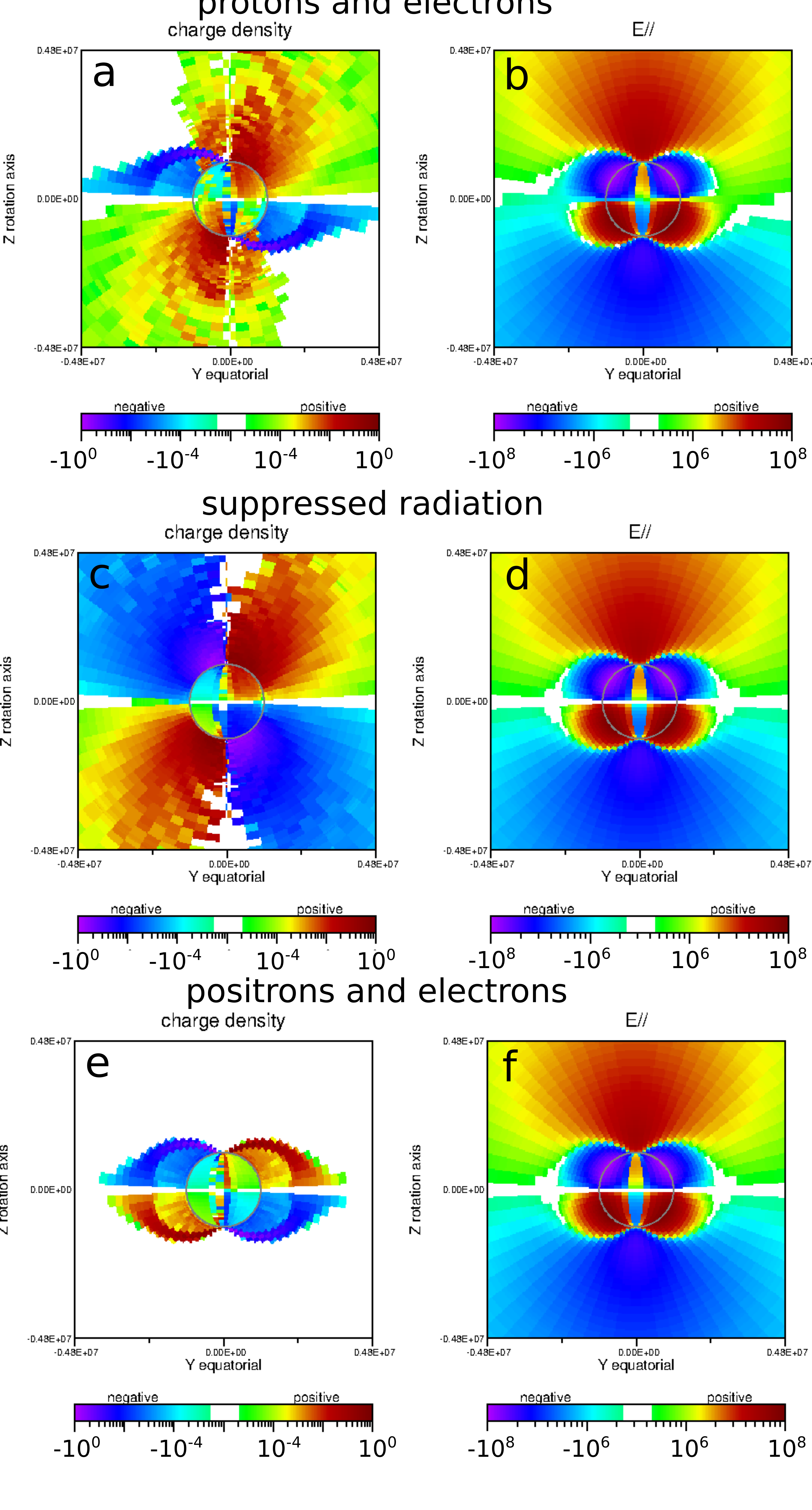}
   \caption{Oblique electrospheres with inclination angle $I=90^o$. The other parameters and the projections are the same as in fig. \ref{fig_aligne_I00_B1E9G_P10ms}. Left-hand side (a,c,e): charge density. Right-hand side (b,d,f) parallel electric field. Upper line (a,b): simulation with electrons and protons and particle radiation loss. Middle line (b,c): electrons and protons and no radiation loss. Lower line (e,f): electrons and positrons and particle radiation losses.  }
              \label{fig_perpendicular_electrosphere_protons_Vigano_protons_sans_FRayt_positons_Vigano_annote}%
\end{figure}

   Three solutions for the perpendicular electrosphere with $i=90^o$ are presented in fig. \ref{fig_perpendicular_electrosphere_protons_Vigano_protons_sans_FRayt_positons_Vigano_annote}. Here again, the plots in the left hand-side column represent the electric charge density, and those in the right hand-side column represent the parallel electric field. On top row (a, b), the electrosphere is simulated with an electrospheric population made of electrons and protons, as in the previously presented simulations. The middle row (c,d) represent the simulation in the same conditions, but where the particle radiation loss are not taken into account. The bottom row shows an electrosphere composed of electrons and positrons. 
   
  Without radiation losses, the regions occupied by the electrons and by the ions have similar sizes. The electrons are not really confined in well delimited domes. Moreover, their energies reach Lorentz factors up to a few $10^8$. With radiation losses, mostly efficient with low mass particles (here, electrons), domes are smaller in size, with a clearly defined boundary.
     
\begin{figure}
   \centering

   \includegraphics[width=8cm]{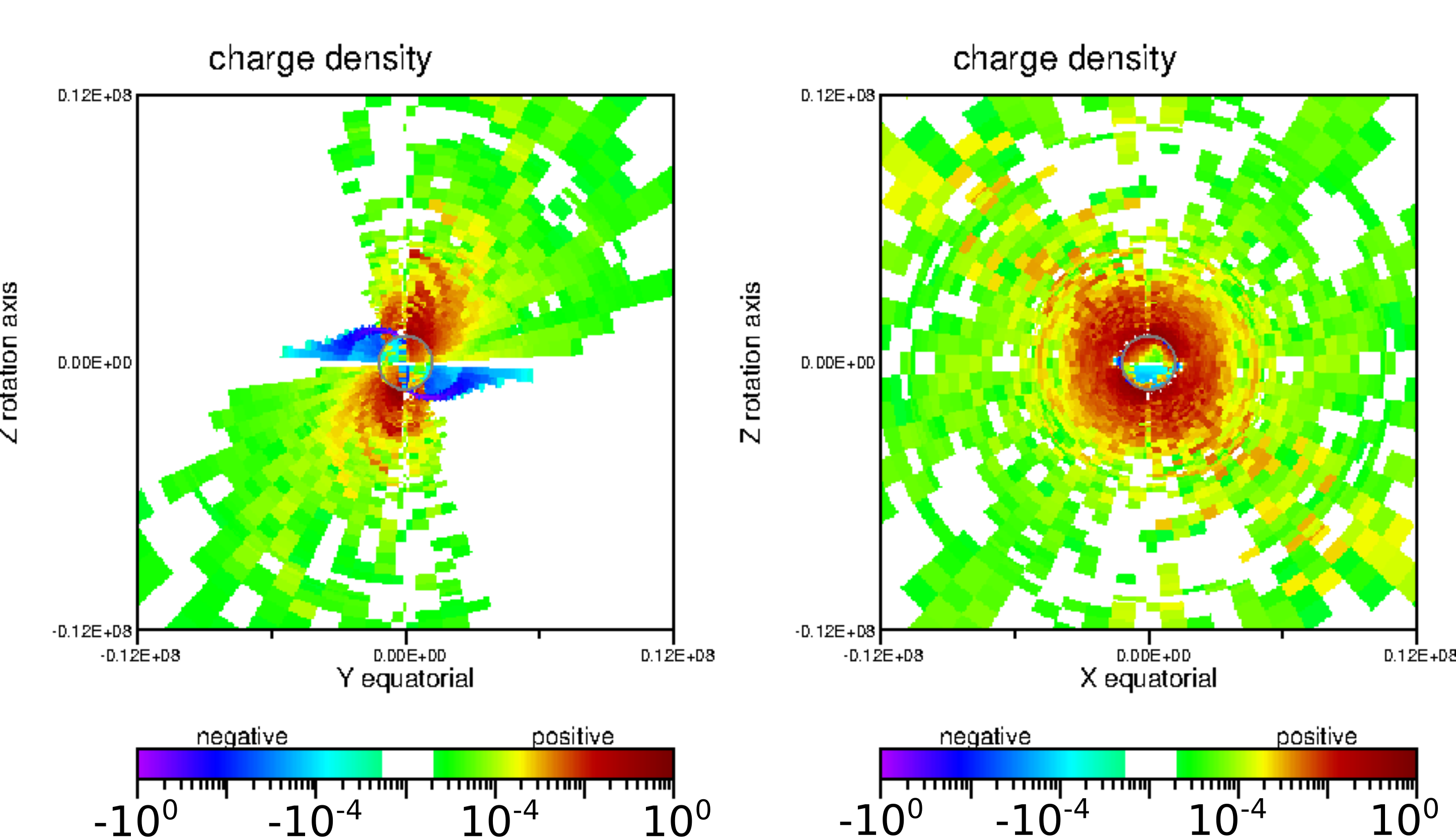}
   \caption{Oblique electrospheres with inclination angle $I=90^o$. The other parameters and the projections are the same as in fig. \ref{fig_aligne_I00_B1E9G_P10ms}. Left-hand side: charge density in the plane containing the rotation axis ($Oz$) and the magnetic axis of symmetry ($Oy$ for $I=90^o$). Right-hand side: charge density in the magnetic equatorial plane ($Oxz$).  }
              \label{fig_perpendicular_electrosphere_proton_torus}%
\end{figure}

   The electrosphere made of electrons and positrons (radiation losses are taken into account) shows a quadrupole structure that was already presented in \cite{McDonald_2009}. This bring the following question : with electrons and proton, is the charge density structure a quadrupole, or constituted of two domes and a proton torus ? In fig. \ref{fig_perpendicular_electrosphere_proton_torus}, the projection of the electric charge density onto the planes $y,z$ and $x,z$ (perpendicular to the magnetic axis for $i=90^o$)  shows a ring of electron all around the magnetic axis and two domes of electrons. With electrons and positrons, there is no particle in the plane $(y,z)$ (not shown). Therefore the dome-torus structure is present for inclination angle $0^o \leq i \leq 90^o$ with electrons and protons. It becomes a quadrupole for electrons and positrons, and $i = 90^o$. 
        
\subsection{Oblique electrospheres in the pulsar graveyard}
\begin{figure}
   \centering

   \includegraphics[width=4cm]{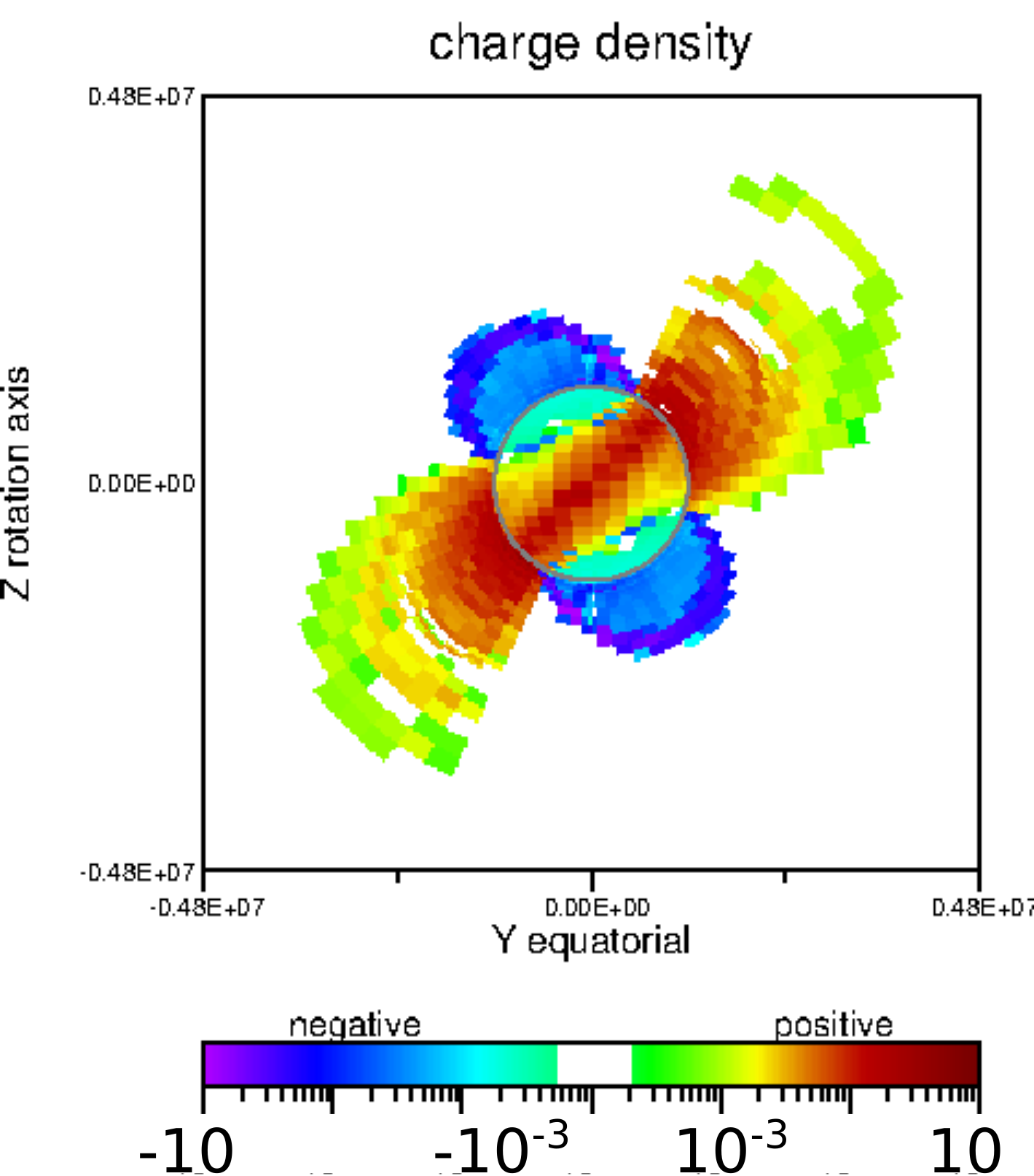}
   \caption{Charge density of an electrosphere in the high magnetic field side of the pulsar graveyard, with $B_*=10^{12}$ G, $P_*=5$ s, $i=45$ degrees.  }
              \label{fig_rho_BE12G_P5s_I45deg}%
    \end{figure}

The electrosphere simulations presented in the previous sections were associated with a NS magnetic field $B_*=10^9$ G, and a rotation period $P=10$ ms. Actually, many recycled pulsars have these characteristics. This means that in these conditions, electron-positron pair productions are sufficient to trigger an intense pulsar radiative activity. 
It is therefore interesting to simulate the electrospheres of NS with lower rotational and magnetic energies, because these ones, situated in the pulsar graveyard, have good chances to behave really as electrospheres, and not as pulsars. 
Therefore, simulations of electrospheres with values of $B_*$ and $P_*$ typical of the pulsar graveyard were performed. Figure \ref{fig_rho_BE12G_P5s_I45deg} shows the electric charge density, with the same projection as in fig. \ref{fig_rho_Eparallel_30_45_75_annote}, of an electrosphere with a surface magnetic field $B_*=10^{12}$ G, a rotation period $P_*=5$ s and an inclination angle $i=45$ degrees. Two  well delimited electron domes and a proton torus are clearly visible. 
 
\begin{figure}
   \centering
   \includegraphics[width=4cm]{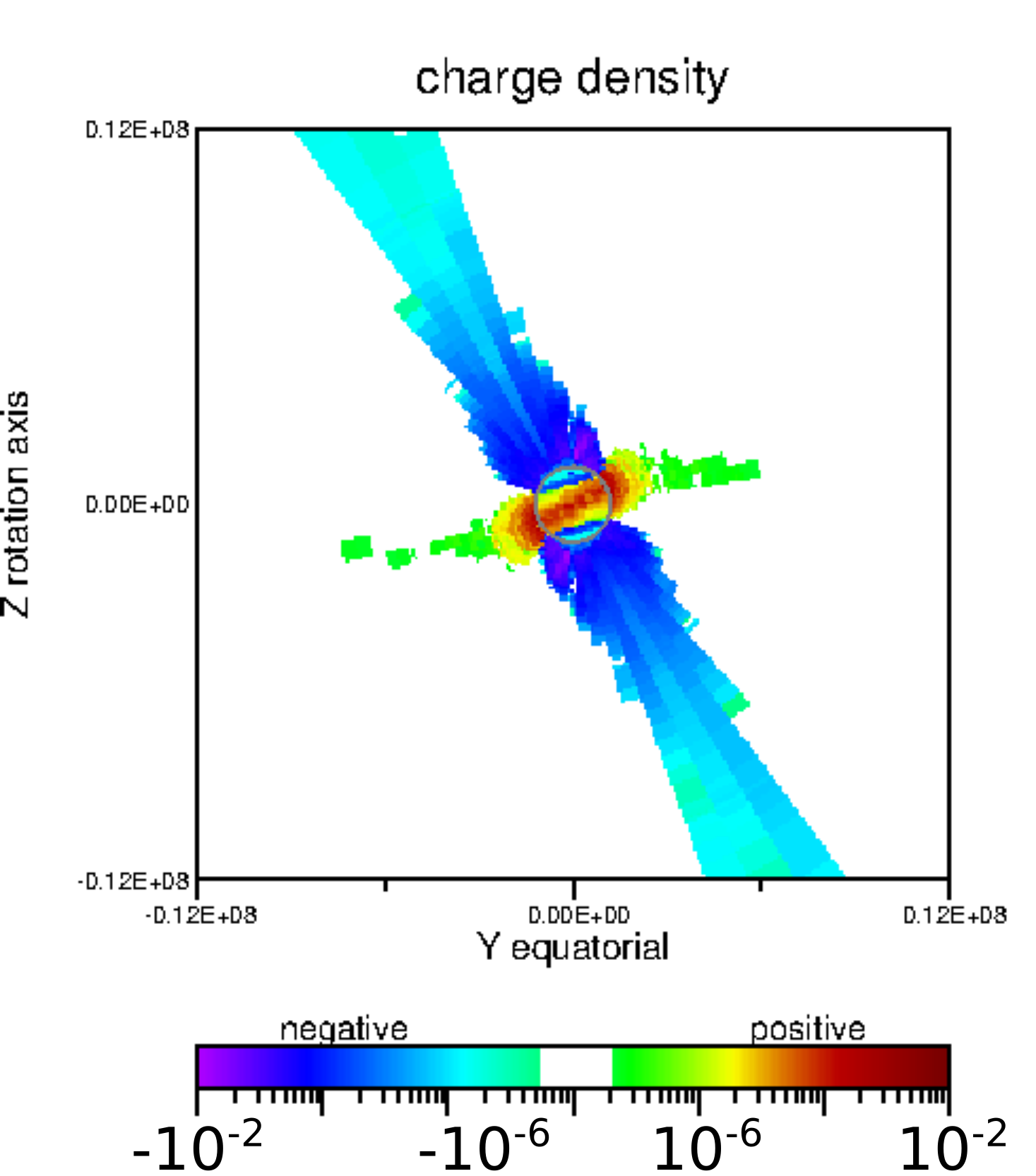}
   \caption{Charge density of an electrosphere in the low magnetic field side of the pulsar graveyard, with $B_*=10^8$ G, $P_*=100$ ms, $i=30$ degrees.  }
              \label{fig_rho_BE8G_P100ms_I30deg}%
    \end{figure}

 Figure \ref{fig_rho_BE8G_P100ms_I30deg} show an example of electrosphere in the low magnetic field area of the $P-\dot P$ diagram. With $B_*=10^8$ G, a rotation period $P_*=100$ ms, and $i=30$ degrees, we still can see the two domes and the proton torus. But the electron domes boundaries extend quite far from the NS. Nevertheless, the areas where the electron density is more that 0.1 \% of its maximal value extends only up to a distance of $4 R_*$. 

\section{Discussion and conclusion} \label{sec_discussion_conclusion}

In this paper, we have presented, for the first time, numerical simulations of oblique electrospheres of neutron stars with realistic parameters. Aligned electrospheres had already been modeled, using realistic parameters (such as those in \cite{Petri_2002}), and two simulations of oblique electrospheres were presented with  reduced parameters in one article \citep{McDonald_2009}. The present paper is based on a significantly larger number of numerical simulations. 

The most striking and immediate observation is the persistence, at any inclination angle, of the structure with two electron domes and an ion torus that was described for aligned electrospheres since the seminal work presented in \cite{Krause_1985}. 

Another observation is that most of the previous models of electrospheres supposed that at least the inner boundary, right above the NS, the plasma was in force free equilibrium. The idea behind this hypothesis, as said in \cite{McDonald_2009} was that the low altitude electrospheric plasma would be sufficiently dense to screen the surface "vacuum-like" electric field, and its parallel component would tend to zero. Our simulations take the finite particle inertia into account, and this does not constrain the solution to be force-free. Moreover, unlike \cite{McDonald_2009}, we did not try to reach a solution that would tend to a force-free equilibrium at the NS surface. What we could see is that the plasma density, that is of the order of the Goldreich-Julian density near the NS, is not high enough to screen the accelerating electric field. In consequence, in spite of still having two electron domes and a torus of protons, they are populated by particles that have been strongly accelerated, and that propagate with very large Lorentz factors. The fact that the boundary conditions for oblique electrospheres may be not force-free may explain why only one work, with no follow up, was published on this subject before the present study.

We have showed the existence of non-force-free solutions to electrosphere. We have not proven their uniqueness, and it is possible that other initial set-ups can favor force-free solutions. 

Because of the force-free condition that they imposed on the NS surface, \cite{Petri_2002} noticed that the total electric charge in the electrosphere $Q_t$ is a free parameter. They presented models of aligned electrospheres with  $Q_t= \alpha Q_c$, where  $Q_c=R_*^3 \Omega_* B_*/ 3c$ and $\alpha$ took the values 0, 1, 3. In the present study, $\alpha$ is determined self-consistently. We observed that it converges toward values that are different in every case, of the order of a few $10^{-2}$. The only exception is when the electrosphere contains electrons and positrons, and $i=90$. In that case, charged particles of opposite signs play the same role, and $\alpha$ varies randomly around zero with an dispersion of a few $10^{-3}$ with the ${E}_{r,\rho}^> =0$ inner boundary condition, and of a few $10^{-6}$ with ${E}_{r,\rho}^> \ne 0$.

The numerical methods involved in the present study are different from those used in the previous numerical studies of electrospheres and pulsars. Here are some comments on their robustness and weaknesses.

This code produces, by construction, a stationary solution. If the actual behavior of a NS environment is oscillatory, it is possible that the solutions we present are a kind of average value of the various intermediate states. However, observation of pulsars (which, in addition to the simulations in this study, contain cascades of electron-positron pairs) shows that all pulses can be different, but that an average of a few tens of pulses in most cases presents a fairly invariant pattern. 
From the point of view of numerical analysis, we have observed that all the parameter sets tested in the present study achieve convergence, but we do not have a general proof of convergence.  

Maxwell's equations are solved with spectral methods. These methods are very efficient as long as no sharp gradient  are met in the simulations (such as spatially unresolved shock waves). With $N_{angles}=32$, the angular resolution may seem fairly low. Simulations conducted with  $N_{angles}=64$ presented smoother solutions for a cost in memory and in CPU multiplied by 4. Nevertheless, the structure and the properties of the electrospheres computed with $N_{angles}=32$ and with  $N_{angles}=64$ were the same; this is why we used $N_{angles}=32$ in most simulations. 

Because of the very high acceleration very near the NS surface, we used a resolution in $r$ down to about 1 cm for the first cells. The  $N_{d} -1$ other shells were tried with various sizes, with $N_{d}$ in the range 8 to 11.  The stability of the solution, with different simulation grids confirmed the robustness of the spectral methods that we implemented.

The integration of long particle trajectories over one iteration is also a factor of efficiency. When a trajectory is long (we used $N_{steps}=10,000$), it can start from the NS neutron star, go across the dome of the proton belt, and go back into the NS star (at a different place from where it started), there is obviously no need to restart it somewhere in the simulation domain at the following iteration. When trajectories are shorter, for instance  with  $N_{steps}=100$, for most of them, the particle is still in the electrosphere when time step 100 is reached, and at the next iteration, it is restarted from inside the electrosphere. It turns out that it is more costly. The solution of convergence is the same, but it is reached at a lower expense when $N_{steps}=10,000$, and with a lesser number of iterations. 

The formalism of charge deposition presented in sec. \ref{sec_particle_transport} assigns each macro-particle a statistical weight proportional to the fraction $f$ of the distribution it represents. Therefore, not all particles have the same statistical weight. This allows to take into account minor populations, and this allows for a wide range of particle number densities, as we could see on plots of the electric charge density. 

The integration of the particle motion is quite robust. It takes finite particle inertia into account. A series of tests were comparison with theoretical well-known solutions is possible were conducted with success. This method of resolution is also very fast, because of the variable time steps $\Delta t_i$ and because there is no need to resolve the particle gyro-period. As said in sec. \ref{sec_motion}, the domain of application of this method is larger than those of pulsars and electrospheres, and it will be presented in a further publication. 

With the simplified model presented in \cite{Higgins_1997}, (\citeyear{Higgins_1998}) it was shown that electrosphere gamma-ray light curves present similarities with those of gamma-ray pulsars. In the present work, the consideration of radiation effects is still poorly taken into account. We do not yet provide maps of radiated photons. This is  a matter of time, and of work to be done. However, the particle energy losses by radiation is taken into account, as described in sec. \ref{sec_radiative}. The method described in this section presents two drawbacks worth mentioning. The first is that the curvature $C_m$ that we should use ideally is those of the trajectory, but, for easiness, we used the curvature of the magnetic field line passing through the particle. Because of the guiding-center drifts, this is not fully correct. The second drawback may have more important consequences. Sometimes (especially when $B_* \sim 10^{12}$), because of the $\gamma^4$ factor in the radiation losses formula, and because of the large values of $\gamma$, a numerical instability can be triggered. As mentioned in section \ref{sec_radiative}, this instability is explosive and easy to detect. When we could detect it, we replaced the general computation of the particle Lorentz factor by a value of $\gamma_a$ given in Eq. \ref{eq_gamma_saturation} that corresponds to an equality of  the electric acceleration and the radiation loss. This does not necessarily correspond to the actual physical situation. The  consideration of the radiation effects can be improved. Also, if it is clear that electrospheres contain high energy particles, the values of the Lorentz factors reached in the simulation must be considered with care. 

The simulations were all conducted on a laptop equipped with an Intel Core i7 processor. The code was parallelized in a non-scalable way. Four processes were used. The typical duration of an iteration of the simulations presented in this paper is typically a few tens of minutes, convergence is reached in about 2 to 4 iterations, and computing 10 iteration requires less than four (clock-wall) hour. 

The low cost of these simulations, in terms of equipment, time and environmental resources, opens up prospects for future developments to improve our understanding of pulsar physics. The next step will involve high-energy photon emissions from electrospheres. Then we plan the inclusion of cascades of electron-positron pairs , and hence numerical simulations of pulsars, if possible with realistic parameters. 
 Many ways of modeling  pair creation have been implemented in numerical simulations. We could for instance implement "uniform particle injection", where pairs are supposed to be created in all acceleration regions. We could also try to give a better account of the micro-physics of pair creations,  if possible by implementing the main features demonstrated in \cite{Voisin_2018a}. 
 The structure of \textit{Pulsar ARoMa} makes it possible to include the metrics of general relativity. The \textit{Pulsar ARoMa}  numerical simulation code is designed to become a versatile tool for the study of pulsar physics. Before reaching this stage, further development and validation tests are necessary, and they are underway.
 
\begin{acknowledgements}
I would like to thank my colleagues Silvano Bonazzola (now retired) and Guillaume Voisin for their help and support.  
\end{acknowledgements}

\bibliographystyle{aa} 

\begin{thebibliography}{25}
\expandafter\ifx\csname natexlab\endcsname\relax\def\natexlab#1{#1}\fi

\bibitem[{{Beskin} \& {Litvinov}(2022)}]{Beskin_2022}
{Beskin}, V.~S. \& {Litvinov}, P.~E. 2022, \mnras, 510, 2572

\bibitem[{{Bonazzola} {et~al.}(2015){Bonazzola}, {Mottez}, \&
  {Heyvaerts}}]{Mottez_2015c}
{Bonazzola}, S., {Mottez}, F., \& {Heyvaerts}, J. 2015, \aap, 573, A51

\bibitem[{{Cerutti} \& {Beloborodov}(2017)}]{Cerutti_2017}
{Cerutti}, B. \& {Beloborodov}, A.~M. 2017, \ssr, 207, 111

\bibitem[{{Deutsch}(1955)}]{Deutsch_1955}
{Deutsch}, A.~J. 1955, Annales d'Astrophysique, 18, 1

\bibitem[{{Gold}(1968)}]{Gold68}
{Gold}, T. 1968, Nature, 218, 731

\bibitem[{{Goldreich} \& {Julian}(1969)}]{Goldreich69}
{Goldreich}, P. \& {Julian}, W.~H. 1969, ApJ, 157, 869

\bibitem[{{Grandcl{\'e}ment} \& {Novak}(2009)}]{Grandclement_2009}
{Grandcl{\'e}ment}, P. \& {Novak}, J. 2009, Living Reviews in Relativity, 12, 1

\bibitem[{{Hewish} {et~al.}(1969){Hewish}, {Bell}, {Pilkington}, {Scott}, \&
  {Collins}}]{Hewish_1969}
{Hewish}, A., {Bell}, S.~J., {Pilkington}, J.~D.~H., {Scott}, P.~F., \&
  {Collins}, R.~A. 1969, \nat, 224, 472

\bibitem[{{Higgins} \& {Henriksen}(1997)}]{Higgins_1997}
{Higgins}, M.~G. \& {Henriksen}, R.~N. 1997, \mnras, 292, 934

\bibitem[{{Higgins} \& {Henriksen}(1998)}]{Higgins_1998}
{Higgins}, M.~G. \& {Henriksen}, R.~N. 1998, \mnras, 295, 188

\bibitem[{{Krause-Polstorff} \& {Michel}(1985)}]{Krause_1985}
{Krause-Polstorff}, J. \& {Michel}, F.~C. 1985, \aap, 144, 72

\bibitem[{{McDonald} \& {Shearer}(2009)}]{McDonald_2009}
{McDonald}, J. \& {Shearer}, A. 2009, \apj, 690, 13

\bibitem[{{Mottez}(2008)}]{mottez_2008_a}
{Mottez}, F. 2008, Journal of Computational Physics, 227, 3260

\bibitem[{{Mottez} {et~al.}(1998){Mottez}, {Adam}, \& {Heron}}]{Mottez_1998}
{Mottez}, F., {Adam}, J.~C., \& {Heron}, A. 1998, Computer Physics
  Communications, 113, 109

\bibitem[{{Pacini}(1967)}]{Pacini67}
{Pacini}, F. 1967, Nature, 216, 567

\bibitem[{{P{\'e}tri}(2013)}]{Petri_2013}
{P{\'e}tri}, J. 2013, \mnras, 433, 986

\bibitem[{{P{\'e}tri}(2015)}]{Petri_2015_multipole}
{P{\'e}tri}, J. 2015, \mnras, 450, 714

\bibitem[{{P{\'e}tri} {et~al.}(2002){P{\'e}tri}, {Heyvaerts}, \&
  {Bonazzola}}]{Petri_2002}
{P{\'e}tri}, J., {Heyvaerts}, J., \& {Bonazzola}, S. 2002, \aap, 384, 414

\bibitem[{{Philippov} \& {Spitkovsky}(2014)}]{Philippov_2014}
{Philippov}, A.~A. \& {Spitkovsky}, A. 2014, \apjl, 785, L33

\bibitem[{{Riley} {et~al.}(2021){Riley}, {Watts}, {Ray}, {Bogdanov}, {Guillot},
  {Morsink}, {Bilous}, {Arzoumanian}, {Choudhury}, {Deneva}, {Gendreau},
  {Harding}, {Ho}, {Lattimer}, {Loewenstein}, {Ludlam}, {Markwardt}, {Okajima},
  {Prescod-Weinstein}, {Remillard}, {Wolff}, {Fonseca}, {Cromartie}, {Kerr},
  {Pennucci}, {Parthasarathy}, {Ransom}, {Stairs}, {Guillemot}, \&
  {Cognard}}]{Riley_Nicer_2021}
{Riley}, T.~E., {Watts}, A.~L., {Ray}, P.~S., {et~al.} 2021, \apjl, 918, L27

\bibitem[{{Smith} {et~al.}(2001){Smith}, {Michel}, \& {Thacker}}]{Smith_2001}
{Smith}, I.~A., {Michel}, F.~C., \& {Thacker}, P.~D. 2001, \mnras, 322, 209

\bibitem[{{Sturrock}(1970)}]{Sturrock_1970}
{Sturrock}, P.~A. 1970, \nat, 227, 465

\bibitem[{{Timokhin} \& {Harding}(2019)}]{Timokhin_2019}
{Timokhin}, A.~N. \& {Harding}, A.~K. 2019, \apj, 871, 12

\bibitem[{{Vay}(2008)}]{Vay_2008}
{Vay}, J.~L. 2008, Physics of Plasmas, 15, 056701

\bibitem[{{Voisin} {et~al.}(2018){Voisin}, {Mottez}, \&
  {Bonazzola}}]{Voisin_2018a}
{Voisin}, G., {Mottez}, F., \& {Bonazzola}, S. 2018, \mnras, 474, 1436

\end{thebibliography}

\end{document}